\documentclass[aps,twocolumn,showpacs,unsortedaddress,amsmath,amssymb,10pt]{revtex4-1}
\usepackage[utf8x]{inputenc}
\usepackage{amsmath}
\usepackage{amsfonts}
\usepackage{bbm}               
\pdfoutput=1
\usepackage{graphicx}
\usepackage{psfrag}
\usepackage[hyperfootnotes=false,pdfpagelabels,pagebackref,unicode]{hyperref}
\usepackage{subfigure}
\usepackage{relsize}
\usepackage{placeins}
 
\graphicspath{{gfx/}}

\newcommand{\InvMT}[1]{\int \limits_{#1 - i \infty}^{#1 + i \infty} \hspace{-8pt}}

\newcommand*\pFqskip{8mu}
\catcode`,\active
\newcommand*\pFq{\begingroup
        \catcode`\,\active
        \def ,{\mskip\pFqskip\relax}%
        \dopFq
}
\catcode`\,12
\def\dopFq#1#2#3#4#5{%
        {}_{#1}F_{#2}\biggl(\genfrac..{0pt}{}{\scriptscriptstyle #3 \displaystyle}{\scriptscriptstyle #4 \displaystyle};#5\hspace{-2pt}\biggr)
        \endgroup
}

\catcode`,\active

\catcode`\,12

\makeatletter
\def\tagform@#1{\maketag@@@{\ignorespaces#1\unskip\@@italiccorr}}
\let\orgtheequation\theequation
\def\theequation{(\orgtheequation)}
\makeatother

\let\orgautoref\autoref
\renewcommand{\autoref}%
        {%
         \orgautoref}

\begin{document}
\title{A time and spatially resolved quench of the fermionic Hubbard model showing restricted equilibration}
\author{Florian Goth and Fakher F. Assaad}
\affiliation{Institut f\"ur Theoretische Physik und Astrophysik,\\
Universit\"at W\"urzburg, Am Hubland, D-97074 W\"urzburg, Germany}
\date{August 12, 2011}

\begin{abstract}
We investigate
the quench of half-filled 1D and 2D fermionic Hubbard models to models without 
Coulomb interaction.
Since the time propagation is gaussian we can use a variety of time-dependent quantum 
Monte Carlo methods to tackle this problem without generating a dynamical
sign problem.
Using a continuous time quantum Monte Carlo method (CTQMC) we achieve a system size
of $128$ sites in 1D, and using a Blankenbecler-Scalapino-Sugar (BSS) type algorithm
we were able to simulate $20 \times 20$ square lattices. Applying these methods 
to study the dynamics
after the quench,
we observe that the final state of the system can be reasonably well described by a 
thermal single-particle density matrix that takes the initial single particle 
conservation laws into account. The characteristic decay towards this limit is found
to be oscillatory with an additional power law decay that depends on the
dimensionality.
This  numerically exact result is shown to compare favorable to mean-field
approximations as well as to perturbation theory.
Furthermore we observe the information propagation in the 1D-case
in the charge charge and spin spin correlations
and find that it is linear with a velocity of roughly $v \approx 4$ in units of the
hopping amplitude.

\end{abstract}

\pacs{71.10.Fd, 05.10.Ln, 05.70.Ln, 73.63.Kv}

\maketitle

\section{Introduction}
The realization of various solid state Hamiltonians using ultra-cold atom experiments has sparked great interest in their out-of-equilibrium physics.
Ongoing experimental work has achieved great control in the preparation of those systems and experimental physicists have started studying 
the dynamics of these systems when they e.g. change parameters of their trapping devices \cite{Trotzky2008, Kinoshita2006}.
In addition, pump and probe femtosecond spectroscopy permits the study of electron relaxation dynamics \cite{PhysRevLett.91.057401, 2009arXiv0910.3808W}.
In this article we propose quantum Monte Carlo methods (QMC) that allow the numerical real-time 
evolution of quantum systems which are prepared in an initial thermal state that contains arbitrary correlations. 
The initial thermal state is computed within the QMC algorithms, which are formulated in terms of a path integral on the Matsubara-Keldysh contour. 
Of special importance is the fact that these methods allow for quenches to arbitrary non-interacting models without introducing
an additional dynamical sign problem.
In this paper we apply these methods to 1- and 2-dimensional half-filled Hubbard models prepared in a thermal initial state and quench them to a Hamiltonian without the Hubbard interaction, that is $U(t>0) = 0$. 
Sotiriadis \cite{SotiriadisCalabreseCardy2009} called this situation with a thermal
initial density matrix a thermal quantum quench in contrast to the pure quantum quench of a pure initial state.

In this setup, we can ask a number of questions concerning the evolution of the system after the quench.
\begin{itemize}
\item Does the system evolve to a new steady-state?
\item How does an isolated system approach a possible new equilibrium?
\item What is the nature of this state?
\item Does the system retain memory of the initial state?
\end{itemize}
In 1D we use an  extension of Rubtsov's CTQMC method \cite{Rubtsov2005, AssaadLang2007}, and in 2D an extended BSS type algorithm \cite{Assaad08_rev}.
A complementary model to ours, where, starting from the free electron limit, the Hubbard interaction was switched on at $t=0$,
was studied theoretically in \cite{PhysRevA.80.061602} and numerically by Kollar et. al. \cite{EcksteinKollar2009} using
dynamical mean-field theory (DMFT).
Kollar et. al. found a critical value of $U_C \approx 3.3$ where the characteristic oscillations in the double occupancy and the fermi surface discontinuity seem to have been suppressed.
They call this a dynamical phase transition as they observe a very fast thermalization in this regime.
More general results for the quench dynamics of a quantum system in arbitrary dimensions have been presented by Moeckel and Kehrein \cite{Moeckel2009}.
Manmana et. al. \cite{Manmana2009} studied a similar problem as ours using time-dependent 
density matrix renormalization group (DMRG) techniques
but in contrast to our spinful electrons
they considered the case of spin-less electrons. They quenched from interaction parameters lying in the metallic or insulating regime to specific values of final Hubbard $U$'s, where they also crossed those phases. They found that the information propagation in their system, as observed in their density density correlation functions, happens only with a finite velocity that depends on the final interaction 
and not instantaneously. 
We also observe this finite velocity of propagation in the charge charge correlation functions for spinful fermions, which gives rise to the notion of a light cone like evolution of the information propagation.
For a number of models this finite velocity of the propagation of information is known as the Lieb-Robinson bound
and was first discovered for quantum spin systems \cite{LiebRobinson1972}. Lieb and Robinson proved that in their system only exponentially small corrections exist outside this light cone. Over the years these theorems got extended
to more systems up to arbitrary harmonic systems on general lattices with local dynamics (see Ref. \cite{CramerSerafiniEisert2008} and references therein). This light cone like structure is a direct consequence of the locality of the dynamics.

The structure of the article and our main results are the following. 
We first define the Hamiltonian and its symmetries in \autoref{sec:model}
and then we carry out mean-field and perturbative calculations to gain insight into the physics at hand.
This is summarized in \autoref{sec:approximations}.
\autoref{sec:algos} describes the two QMC algorithms used (CTQMC and BSS) to study the physics of the quench, in a self-contained manner.
Our exact numerical results are presented in \autoref{sec:numresults}
and compare favorably with the analytic calculations.
We find that  in 1D and  2D local quantities   equilibrate to values that can be reasonably well described by an effective single particle density matrix that respects the particle densities $n_k$ of the initial thermal density matrix. The approach to the equilibrium follows a  dimension dependent  power law. 
For single particle quantities  such as Green functions  this power law follows that of a diffusion process, $t^{-D/2}$,  where $t$ is the time. 
Finally we take a look at the information propagation of correlations in real space through the system.
We show that a light cone like structure exists, beyond which the propagation of the information of the correlation is exponentially suppressed.
Details of the calculations are presented in appendices.  
\section{The model and its symmetries}
\label{sec:model}
Using the well-known fermionic operators where $c^\dagger_{i\sigma}$ creates an electron
and $c_{i\sigma}$ annihilates an electron
we can define the Hubbard model:
\begin{equation}
 H = \underbrace{- \sum \limits_{ij \sigma} t_{ij} c^\dagger_{i\sigma} c_{j\sigma}}_{=:H_0} + \underbrace{U \sum \limits_i (n_{i\uparrow} - \frac{1}{2})(n_{i\downarrow} - \frac{1}{2})}_{=:H_U}.
 \label{eq:Hubbard-model}
\end{equation}
Here $n_{i\sigma} = c^\dagger_{i\sigma}c_{i\sigma}$.
For our simulations, we restrict ourselves to the case of nearest-neighbor hopping on hyper-cubic
lattices with the lattice constant set to unity.
\begin{equation}
 t_{ij} = 
 \begin{cases}
  1 & \text{if $i$, $j$ are nearest neighbors}\\
  0 & \text{otherwise}
 \end{cases}
\end{equation}
and we restrict ourselves to half-band filling. 
With this choice of parameters, the negative sign problem does not plague the evaluation of the initial density matrix. 
Defining the hopping matrix in that way we set the energy unit to the amplitude of the hopping. 
This frees the letter $t$ and enables us to use it for denoting the real time $t$.
The above defines our unnormalized initial density matrix:
\begin{equation}
 \rho(t=0) = e^{-\beta H}
\end{equation}
describing a  Mott insulating state at inverse temperature $\beta$. At $t=0$ we switch off the Hubbard interaction, that is $H_U(t>0) = 0$, such that the unitary time evolution of the system 
is given by
\begin{equation}
 U(t, t') = e^{-iH_0 (t-t')}.
\end{equation}
As the evolution of the system is carried out with the hopping Hamiltonian $H_0$, there are a number of conserved quantities.
In particular the $k$-space resolved particle density $n_{k\sigma}$ and all related quantities, such as the kinetic energy, are conserved
since $[n_{k\sigma}, H_0] = 0$.


In addition, at the particle-hole symmetric point, where
\begin{equation}
 \epsilon(\vec{k}) = -\epsilon(\vec{k}-\vec{Q})
\end{equation}
holds, $\eta$-pairing
\begin{equation}
\eta_{\vec{Q}}^\dagger = \sum \limits_{\vec{k}} c^\dagger_{\vec{k},\uparrow} c^\dagger_{-\vec{k}+\vec{Q},\downarrow}.
\end{equation}
is a conserved quantity, since
\begin{equation}
 \eta^\dagger_{\vec{Q}}(t) = \sum \limits_{\vec{k}} e^{it(\epsilon(\vec{k}) + \epsilon(-\vec{k} + \vec{Q}))} c^\dagger_{\vec{k}\uparrow}c^\dagger_{-\vec{k}+\vec{Q},\downarrow}
\end{equation}
with $\epsilon(\vec{k}) = -2 \sum \limits_{i=1} ^D\cos(k_i)$ in $D$ dimensions and $\vec{Q} = \pi \sum \limits_{i=1}^D \vec{e}_i$ where $\vec{e}_i$ denotes cartesian unit vectors.
\begin{figure}
 \includegraphics[width=\linewidth]{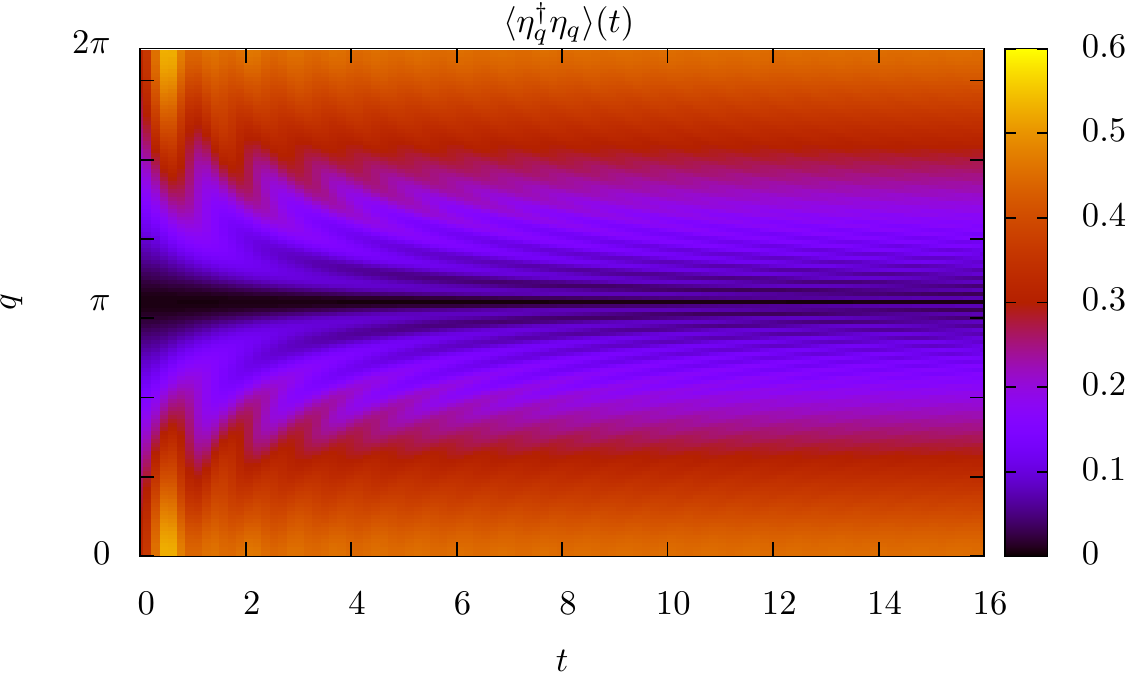}
 \caption{The momentum resolved $\eta$ - pairing correlation function as a function of $t$ for an initial Hubbard interaction of $U=2$ on a $L=128$ site chain. One can clearly see the horizontal black line in the middle of the figure which is present for all times. This corresponds to the fact that this special value of $q = \pi$ is a conserved quantity.}
 \label{fig:Eta_kSpace_U2}
\end{figure}
Therefore $\eta$-pairing provides a non-trivial test of the QMC methods.
\autoref{fig:Eta_kSpace_U2} shows the result of a Monte Carlo run. The black line that is present for all times, in the middle of the figure
corresponds to the particle-hole symmetric point in the $\eta$-pairing. Its existence provides a non-trivial test for the QMC methods
that we outline in \autoref{sec:algos}.
\section{Mean-field analysis and Perturbation theory}
\label{sec:approximations}
\subsection{Mean-field analysis}
\begin{figure}[ht]
\begin{center}
 Analysis of mean-field magnetization
\end{center}
\vspace{-0.2cm}
 \includegraphics[width=\linewidth]{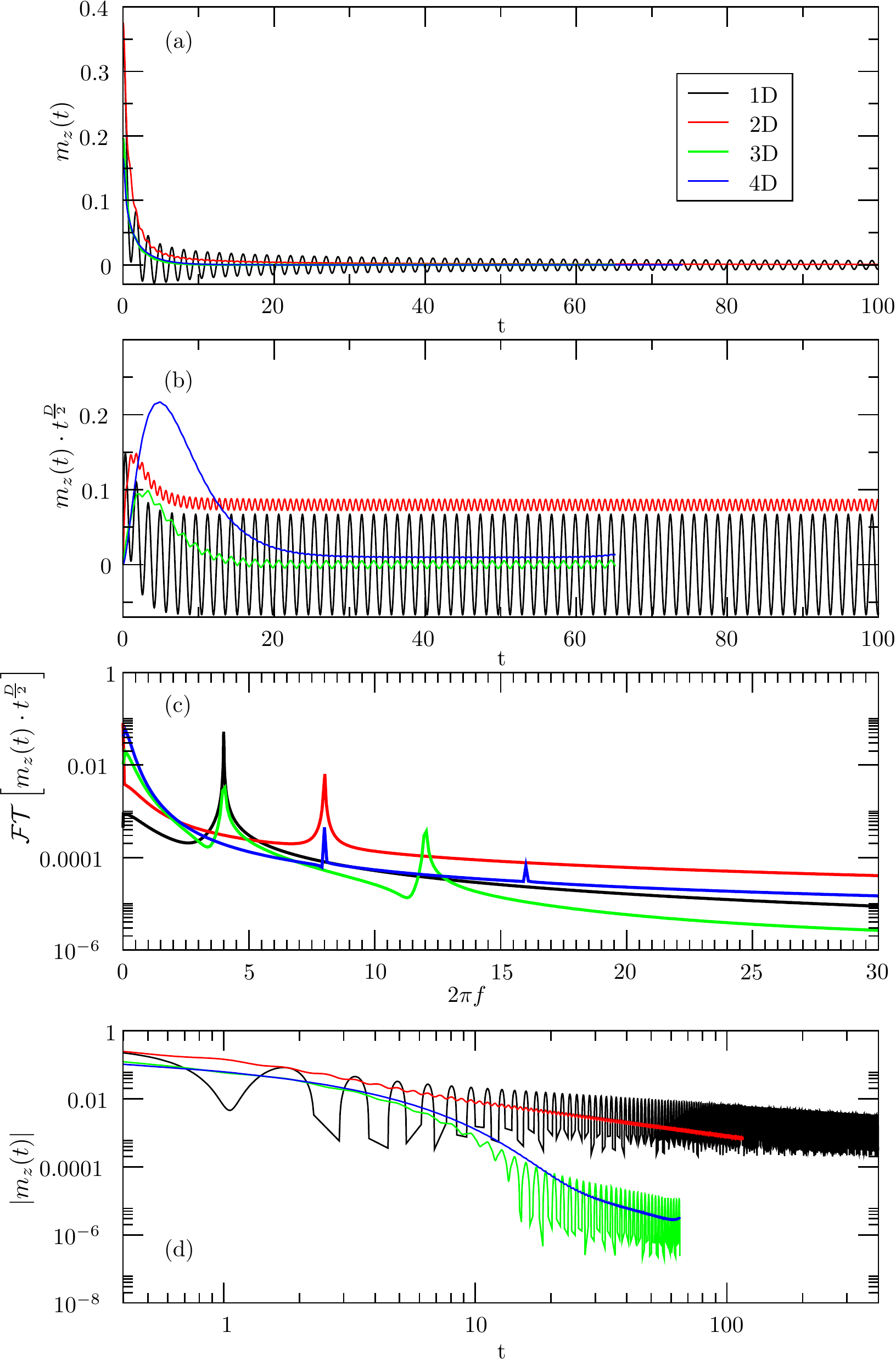}
\caption{The time-dependent decay of the magnetization of the mean-field solution with $T=0$ and $U=2$.
(a) shows the behaviour of \autoref{eq:mztabstract} in dimensions from 1 to 4.
The equations for the magnetization were solved self-consistently and afterwards the time-propagation was calculated.
The colors are consistent through all four plots.
Now in (b) every data set is multiplied by the expected $t^{\frac{D}{2}}$ behaviour.
We see that all data sets approach sines with constant amplitude, thereby 
confirming our conjecture \autoref{eq:suspicion} without introducing the numerical artifacts of the log-log-plot (d).
(c) is the Fourier transform of the data sets in (b).
Finally (d) is the double logarithmic plot of the absolute value of the data sets in (a).
The kinks in the data for 1D and 3D are due to the fact that we have data with a very fine time-resolution.
The oscillations in these graphs would extend all the way to zero in the plot, thereby effectively coloring the plot in black.
Nevertheless the power-law decay of the envelopes is clearly visible.}
\label{fig:mz_log_log}
\end{figure}

We carried out a mean-field approximation based on an anti-ferromagnetic decomposition
of the Hubbard Hamiltonian.
In 1D, the unit-vector $a_1 = \left(1\right)$, and in 2D, the basis $a_1 = \left( 1,0\right)^T$ and $a_2 = \left( 0,1\right)^T$
span the lattice.
To have the possibility of an anti-ferromagnetic ordering in the mean-field approximation we define the anti-ferromagnetic unit cell in 1D simply as twice as 
large as $A_1 = (2)$ and in 2D we define the unit-vectors as $A_1 = (1,1)^T$ and $A_2 = (1,-1)^T$. The unit cell now contains two orbitals and we 
label one of them with $c_i$ and the other with $d_i$.
We define the Fourier transform of these operators as
\begin{equation}
 \begin{split}
  c_R &= \frac{1}{\sqrt{N}} \sum \limits_k e^{ikR} c_k\\
  d_R &= \frac{1}{\sqrt{N}} \sum \limits_k e^{ikR} d_k\\
 \end{split}
\end{equation}
where $R$ denotes lattice vectors labeling anti-ferromagnetic unit cells.
With the definition of the anti-ferromagnetic unit cell and the labeling of its electrons we 
can now proceed to describe the calculation in a unified way.
Additionally we introduce $n^d_{R \sigma} = d^\dagger_{R\sigma} d_{R\sigma}$
as well as $n^c_{R\sigma} = c^\dagger_{R\sigma} c_{R\sigma}$ as shorthand notation.
We rewrite the Hubbard interaction as
\begin{equation}
 H_U = \frac{-U}{2}\sum \limits_R \left[\left(n^c_{R\uparrow} - n^c_{R\downarrow}\right)^2
 + \left(n^d_{R\uparrow} - n^d_{R\downarrow}\right)^2 \right].
\end{equation}
In this mean-field approximation we introduce the mean-field order parameter $m_z$:
\begin{equation}
 \begin{split}
  m_z &= \left<n^c_{R\uparrow} - n^c_{R\downarrow}\right>\\
 -m_z &= \left<n^d_{R\uparrow} - n^d_{R\downarrow}\right>.
 \end{split}
\end{equation}
such that the mean-field Hubbard interaction reads
\begin{equation}
 H_U^{\text{MF}} = -\frac{U m_z}{2} \sum \limits_R \left(n^c_{R\uparrow} - n^c_{R\downarrow}\right)
 - \left(n^d_{R\uparrow} - n^d_{R\downarrow}\right).
\end{equation}
Carrying out the Fourier transform on the kinetic energy part and introducing the quantity $\gamma_{k\sigma} = \left( c_{k\sigma}, d_{k\sigma}\right)^T$
we can rewrite the total Hamiltonian in a matrix form like 
\begin{equation}
 H = \sum \limits_{k\sigma}
 \left( c_{k\sigma}^\dagger, d_{k\sigma}^\dagger\right)^T
 \underbrace{\left( \begin{array}{cc} -\Delta \sigma& Z_k\\
                           \bar{Z_k} & \Delta \sigma
                          \end{array}
                  \right)}_{=:H_0(k,\sigma)}
             \left( \begin{array}{c} c_{k\sigma} \\
                           d_{k\sigma}
                          \end{array}
                  \right). 
\end{equation}
Here $Z(k)$ contains the information about the lattice structure, $Z(k) = - \left(1+e^{-ikA}\right)$ in 1D and $Z(k) = - \left(1+e^{-ikA_1}\right)\left(1+e^{-ikA_2}\right)$ in 2D.
$\Delta = \frac{m_z U}{2}$ and the band structure is determined by $E(k) = \pm \sqrt{\Delta^2 + |Z(k)|^2}$. This, so far, is well known
from thermodynamics. Next we need to introduce real-time dynamics into this setting.
The Hamiltonian responsible for the time evolution contains no explicit time-dependence, therefore
the evolution of $\gamma$ is given by:
\begin{equation}
 \gamma_{k\sigma}(t) = e^{i H_0 t} \gamma_{k\sigma} e^{-iH_0 t}.
\end{equation}
To obtain a Heisenberg equation of motion, we derive this equation with respect to $t$. Taking into account the structure of $\gamma$ as well as the fermionic commutation rules, we get 
\begin{equation}
 \frac{d}{dt} \gamma_{k\sigma,\alpha}(t) = -i (H_{0}(k,\sigma)\gamma_{k\sigma}^\dagger(t))_\alpha
\end{equation}
This is solved by 
\begin{equation}
 \gamma_{k\sigma}(t) = e^{-iH_0(k,\sigma) t}\gamma_{k\sigma}.
\end{equation}
The time dependence of the magnetization is then given by 
\begin{equation}
 m_z(t) = \frac{1}{N} \sum_{k\sigma} \langle \gamma^\dagger_{k\sigma}(t) \sigma_z \gamma_{k\sigma}(t)\rangle .
 \label{eq:mztabstract}
\end{equation}
This quantity is plotted in \autoref{fig:mz_log_log} (a) as a function of dimension $D$ and at $T=0$.
From the log-log plot (\autoref{fig:mz_log_log} (d)) we clearly see that the maxima of the oscillations can be fitted by functions that decay as a power-law.
To get rid of certain numerical artifacts in the log-log-plot of the absolute value we also plot $m_z(t) \cdot t^{D/2}$ in \autoref{fig:mz_log_log} (b).
We see that this quantity quickly 
approaches a simple sine wave oscillation pattern with constant amplitude. 
In \autoref{fig:mz_log_log} (c) we show the Fourier transform of this quantity.
In 1D only a single frequency of $\omega \approx 4$ is present whereas 2D oscillates with a frequency of $\omega \approx 8$.
Depending on the dimensionality we see an odd-even effect.
Even dimensions have the same base frequency as in $2D$ and odd dimensions have the same base frequency as $1D$ has.
Additionally, higher harmonics show up in $D=3$ and $D=4$ with a spacing of $\Delta \omega \approx 8$.
Judging from this plot we propose that the long-time behaviour of the envelope of this decay is connected with the dimensionality of the system
like 
\begin{equation}
 |m_z(t)| \propto t^{-\frac{D}{2}}.
 \label{eq:suspicion}
\end{equation}
In the thermodynamic limit we can perform a more detailed analysis of the behaviour of the 1D magnetization.
After some calculation we get
\begin{equation}
 m_z(t) = \frac{2\Delta}{\pi} \int \limits_0^2 dx \frac{\cos (2 t x) }{\sqrt{4-x^2}\sqrt{x^2+\Delta^2}}.
\end{equation}
The limit $\Delta \rightarrow \infty$ is exactly solvable and is a representation of the Bessel function $J_0$, therefore
$m_z(t) = J_0(4 t)$.
The leading order asymptotic behaviour of the 1D magnetization with respect to $t$ is
given by 
\begin{equation}
 m_z(\Delta,t) = (1+\frac{4}{\Delta^2})^{-\frac{1}{2}} J_0(4t).
\end{equation}
The calculation is outlined in \autoref{subsec:mag1Dasymptotic}.
The well-known asymptotic behaviour for large $t$ of $J_0$ is
\begin{equation}
 J_0(4 t) = \sin (4 t + \frac{\pi}{4}) \frac{1}{\sqrt{2\pi}} \frac{1}{\sqrt{t}} + O(t^{-3/2}),
\end{equation}
therefore confirming our hypothesis in 1D.
In \autoref{subsec:magnetizationND} we generalized this to arbitrary dimensions and give an asymptotic expansion
of $m_z(t,\Delta)$ with respect to $\Delta$.
A key observation is that in dimension $D$ the limit $\Delta \rightarrow \infty$ is exactly solvable
\begin{equation}
 m_{D}(t) := m_{z,D}(t,\Delta \rightarrow \infty)  \propto J_0^D(4t),
\end{equation}
therefore the asymptotic behaviour to lowest order in time of the magnetization is
\begin{equation}
 m_{D}(t) \propto t^{-\frac{D}{2}} \sin^D(4t + \frac{\pi}{4})
\end{equation}
which fits well to our numerical observations at a finite $\Delta$.
The observed frequencies can be explained by using power reduction formulas for trigonometric functions \cite{GradshteynRyzhik}.
In the following $\omega_{D,k} = 4(D-2k)$ denotes the $k$'th frequency.
For even dimension we have
\begin{equation}
 \sin^D(4t)\hspace{-1pt} = \hspace{-1pt}
 \frac{1}{2^{D}}\hspace{-3pt}
 \left[2\hspace{-4pt}\sum \limits_{k=0}^{\frac{D}{2}-1} (-1)^{\frac{D}{2}-k}\binom{D}{k} \cos(\omega_{D,k}t) +\hspace{-1pt} \binom{D}{\frac{D}{2}}\right].
\end{equation}
Therefore in even dimensions the smallest observed frequency is $2 \omega_0$ where $\omega_0$  is some base frequency(in our case we have $\omega_0 = 4$)
and the largest observed frequency is $D \omega_0$.
To this observation fits a simpler result derived by using a flat band of bandwidth $2w$ in \autoref{subsec:mzconst}.
The behaviour in the limit of $\Delta \rightarrow \infty$ is
\begin{equation}
 m_z^{\text{const}}(w,t) \propto \frac{\sin(2wt)}{2wt}.
\end{equation}
Hence the decay of the system with a constant density of states is similar to a 2D Hubbard system with bandwidth $2w = 8$.
This result is consistent with the $t^{-\frac{D}{2}}$ law, since a constant density of states is realized by free electrons in a 2D continuum.
Here it is obvious that the frequency of the oscillations depends on the bandwidth.
In odd dimensions  the powers of the sine are given by
\begin{equation}
 \sin^{D}(4t) = \frac{1}{2^{D-1}} \sum \limits_{k=0}^{\frac{D-1}{2}} (-1)^{\frac{D-1}{2}+k} \binom{D}{k} \sin(\omega_{D,k}t).
\end{equation}
In odd dimension we conclude that the lowest observable frequency is $\omega_0$ and 
that the largest frequency is again $D \omega_0$.
We see that in the large-$t$ regime the frequencies are given by $\omega_k = (D - 2k)\omega_0$. Therefore it is clear that the difference between two frequencies is 
$\Delta\omega = 2\omega_0$ which gives in our case the observed $\Delta \omega = 8$.
By noting that the squared magnetization is related to the double occupancy by 
\begin{equation}
 \frac{2}{N}\sum \limits_i \langle (n_{i\uparrow} - n_{i\downarrow})^2\rangle = 1- \frac{2}{N}\sum \limits_i \langle n_{i\uparrow} n_{i\downarrow}\rangle
\end{equation}
we expect to see similar behaviour in accessible correlation functions in our QMC data.
We conclude this section by noting that  for a fixed linear dimension 
size effects set in at the same time.  This is consistent with the 
notion of a dimension independent velocity for the propagation of 
information.

\subsection{Perturbation theory}
\begin{figure}
\begin{center}
 Comparing Monte Carlo data and perturbation theory
\end{center}
\psfrag{0.58}[Bl][Bl][0.7]{$0.58$}
\psfrag{0.59}[Bl][Bl][0.7]{$0.59$}
\psfrag{0.6}[Bl][Bl][0.7]{$0.6$}
\psfrag{0.61}[Bl][Bl][0.7]{$0.61$}
\psfrag{0.62}[Bl][Bl][0.7]{$0.62$}
\psfrag{0.63}[Bl][Bl][0.7]{$0.63$}
\psfrag{0.64}[Bl][Bl][0.7]{$0.64$}
\psfrag{0.65}[Bl][Bl][0.7]{$0.65$}

\psfrag{0}[Bl][Bl][0.7]{$0$}
\psfrag{1}[Bl][Bl][0.7]{$1$}
\psfrag{2}[Bl][Bl][0.7]{$2$}
\psfrag{3}[Bl][Bl][0.7]{$3$}
\psfrag{4}[Bl][Bl][0.7]{$4$}
\psfrag{5}[Bl][Bl][0.7]{$5$}
\psfrag{theory}[Bl][Bl][0.8]{$\propto J_0^4(4t)$}
\psfrag{MonteCarloData}[Bl][Bl][0.7]{Monte Carlo Data}
\psfrag{Sofq}[Bl][Bl][0.7]{$S(\vec{q} = (\pi, \pi)^T, t)$}
\includegraphics[width=\linewidth]{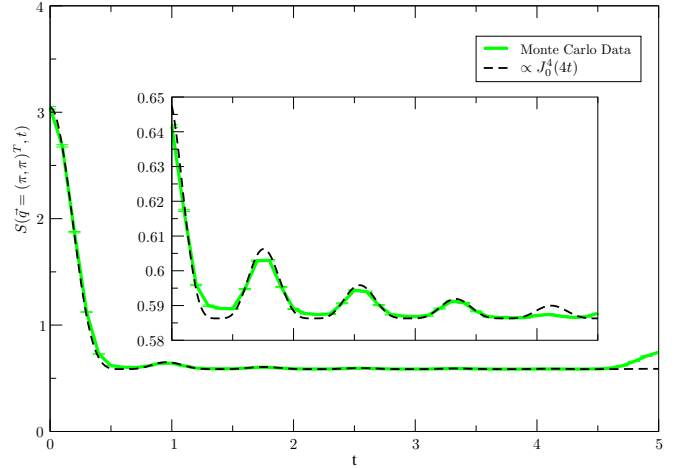}
\caption{The spin spin correlation function $S(\vec{q} = (\pi, \pi)^T, t)$ of a 2D Monte Carlo simulation of a $20 \times 20$ lattice
at $U=8$ and $\beta = 2.5$. We expect this to be in the high-temperature regime where \autoref{eq:Soft_small_beta} is valid.
The dashed black line is $\propto J_0^4(4t)$. The offset is taken from the large time behaviour of the QMC data and the
amplitude was taken from the value at $t=0$. The inset shows a magnified view of the region below the inset from $t=1$ to $t=4.5$.
We see that the approximation works almost \emph{flawlessly} in that regime. The deviation for $t\rightarrow 5$ has its root in boundary effects that set in for approaching that point in time.
}
\label{fig:comparison_to_Bessel_to_the_fourth}
\end{figure}

Here we consider a first order expansion in the strength of the interaction in the thermal Hamiltonian.
This is sufficient since no interaction is present in the real-time evolution.
Additionally, the expansion has the advantage that we can analytically perform the thermodynamic limit.
We consider the spin spin correlation function
\begin{equation}
\begin{split}
 &S(R_i, t)  = \sum \limits_{\sigma \sigma'} \sigma \sigma'  \langle n_{i\sigma}(t) n_{0 \sigma'} (t)\rangle \\
 & = \sum \limits_{\sigma \sigma' kpq} \hspace{-1pt}\frac{\sigma \sigma'}{N^2} e^{i R_i q} \langle c^\dagger_{k\sigma}(t) c_{k-q \sigma}(t) c^\dagger_{p \sigma'}(t) c_{p+q \sigma'}(t) \rangle \\
 & = \hspace{-4pt}\sum \limits_{\sigma \sigma' kpq} \hspace{-4pt} e^{i R_i q} \frac{\sigma \sigma'}{N^2} \langle S_{\sigma \sigma' }(k,p,q,t) \rangle = \frac{1}{N} \sum \limits_q e^{i R_i q} S(q,t).
 \end{split}
\end{equation}
The usual perturbative expansion of $\rho$ gives
\begin{equation}
 \langle S_{\sigma \sigma'}(k,p,q,t) \rangle = \langle (1 - \int \limits_0^\beta d\tau H_U(\tau)) S_{\sigma \sigma'}(k,p,q,t)\rangle_0 .
\end{equation}
This result is the same as obtained by an expansion of the full Keldysh evolution operator $S_C$.
It is obvious that all contributions stem from correlation functions that mix real-time and imaginary-time.
Therefore a solution of this problem using a plain Keldysh method along the real-time contour is not possible.
After Wick-decomposing this expression and collecting the remaining terms, we get for the 
spin spin correlation function $S(q,t)$: 
\begin{equation}
\begin{split}
S(q,t) &= -\frac{2}{N}\sum \limits_{k} \langle n_{k-q}\rangle \langle n_k \rangle - \\
&\frac{2U}{N^2}\int \limits_0^\beta d\tau \sum \limits_{kp} G^<_{p+q}(\tau, t) G_p^>(t,\tau)G_{k-q}^<(\tau, t) G_k^>(t,\tau).
\end{split}
\end{equation}
To interpret the dimensional dependence in QMC simulations we will take a closer look at $S(\pi, t)$, since it is related to the magnetization.
Neglecting the time-independent zeroth order contribution we get:
\begin{eqnarray}
 S(\pi,t) = \text{const.} - 2U \int \limits_0^\beta d\tau \xi(t,\tau) \xi(t, \tau), \\
\begin{aligned}
\xi(t, \tau) &= \frac{1}{N} \sum \limits_k (f(\beta \epsilon_k) - 1)^2 e^{i \epsilon_k(t+i \tau)}\\
 &=\int \limits_{-\infty}^\infty d\epsilon g(\epsilon) (f(\beta \epsilon) - 1)^2 e^{i\epsilon(t + i \tau)}.
 \end{aligned}
\label{eq:xit}
\end{eqnarray}
In the last line of \autoref{eq:xit} we performed the thermodynamic limit and introduced the density of states $g(\epsilon)$. $f(\epsilon)$ denotes the usual Fermi function.
Performing the $\tau$-integral in $S(t)$ and rearranging terms we get:
\begin{equation}
\begin{split}
 S(\pi,t) &\propto \int \limits_{-\infty}^\infty d\epsilon_k\int \limits_{-\infty}^\infty d\epsilon_p \frac{g(\epsilon_p) g(\epsilon_k)}{\epsilon_k + \epsilon_p}e^{2it(\epsilon_p + \epsilon_k)} \times\\
 &\times\left[ f^2(\beta, -\epsilon_k) f^2(\beta, -\epsilon_p) - f^2(\beta, \epsilon_k) f^2(\beta, \epsilon_p)\right].
 \end{split}
 \label{Soft}
\end{equation}
The large bracket that contains all Fermi functions has an expansion in $\beta$ as $\frac{\beta}{8}\cdot(\epsilon_p+\epsilon_k) + O(\beta^3)$.
Therefore, to first order the denominator in \autoref{Soft} is canceled and the two remaining integrals decouple.
Then we have 
\begin{equation}
 S(\pi,t) \propto \left[ \int \limits_{-\infty}^\infty d\epsilon g(\epsilon) e^{2i\epsilon t}\right]^2.
 \label{eq:Soft_small_beta}
\end{equation}
Specializing to the density of states of a 1D chain $g_{1D}(\epsilon) = \frac{\Theta(4-\epsilon^2)}{\sqrt{4-\epsilon^2}}$ we get
$S(\pi, t)\propto J_0^2(4t)$ with $J_0$ the Bessel function of the first kind.
Since the spin spin correlation in the high temperature limit is just the mean-field magnetization squared we can deduce from the general result in \autoref{subsec:magnetizationND} that
the leading order decay of $S(\pi, t)$ is like $S(\pi, t)\propto t^{-D} \sin^{2D}{(4 t)}$. 
\autoref{fig:comparison_to_Bessel_to_the_fourth} shows a comparison of this theoretically predicted behaviour in this approximation to an exact Monte Carlo run in 2D.
We note that the large $\beta$ limit gives the same leading order behaviour.

\section{Description of the QMC Algorithms}
\label{sec:algos}
The physics of the Hubbard model is usually not reasonably well described by the simple approximations of the previous section.
Especially the mean-field description of the 1D Hubbard model is usually just plain wrong
since it fails to describe its low energy Luttinger liquid physics.
Therefore we have to verify and extend our approximate results using unbiased numerical QMC methods.
Starting from the average of an operator $O_H(t)$ in the Heisenberg picture, we have the time-dependent average
\begin{equation}
\begin{split}
 O(t) & = \text{Tr} \left( \rho O_H(t) \right) \\
 & = \text{Tr} \left(\rho U(0,t) O_S(t) U(t, 0) \right)
 \end{split}
 \label{eq:average}
\end{equation}
with the density matrix $\rho$ of the Hubbard model \autoref{eq:Hubbard-model}, the time evolution operator $U(t,t')$
and the possibly explicitly time-dependent operator $O_S(t)$ in the Schr\"odinger picture.
This equation forms the basis of the stochastic methods outlined in this section.
Inserting the identity $\mathbbm{1}$ gives the partition function which allows the construction 
of the Markov chain Monte Carlo method in \autoref{subsec:DDQMCEXP}.
Restricting the real-time dependence in $U(t,t')$ to single particle Hamiltonians enables
the treatment of time-dependent problems by using BSS type algorithms as outlined in \autoref{subsec:AUXQMC}.
\linebreak
\subsection{The weak-coupling CTQMC approach for real-time evolution}
\label{subsec:DDQMCEXP}
Starting from the partition function we derive a weak-coupling CTQMC
 method on the full contour similar as done in Refs. \cite{AssaadLang2007, DavidsPaper2010, Rubtsov2005}
on the imaginary contour.
\begin{figure}
\includegraphics[width=\linewidth]{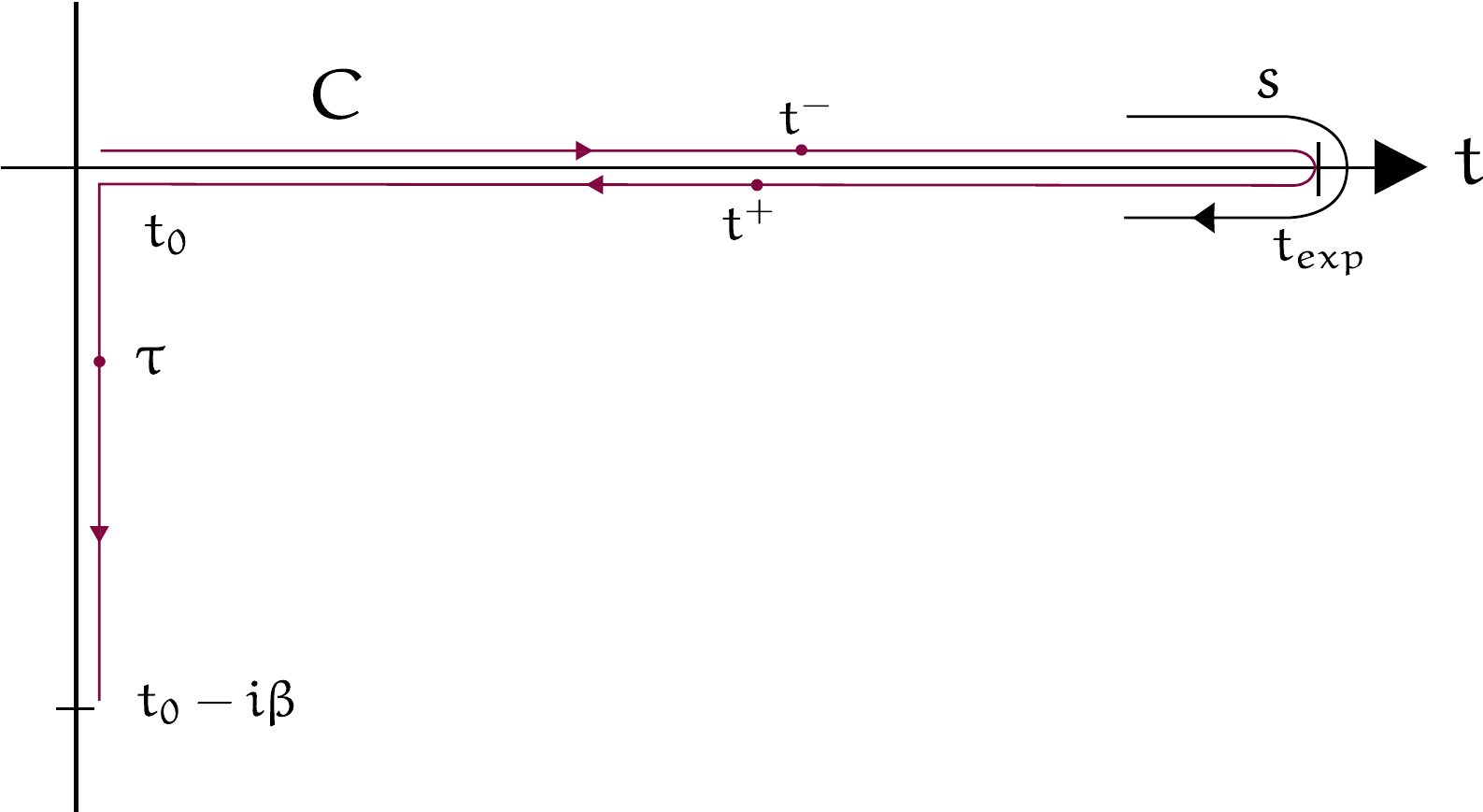}
\caption[The Full Contour]{The full contour $C$ that enables us to cover imaginary-time evolution and real-time evolution on a common footing. $t^-$ is a time on the forward branch, $t^+$ is a time on the backward branch and $\tau$ is a time on the imaginary branch. This contour is parametrized by the contour-time $s$, that runs from $0$ to $t_{exp}$ on the forward contour, from $t_{exp}$ to $2 t_{exp}$ on the backward contour and from $2 t_{exp}$ to $2 t_{exp} + \beta$ on the imaginary branch.
Note that this mapping of imaginary- and real-time into the complex plane leads imaginary time (the one familiar from thermodynamic perturbation theory) to end up on the imaginary axis, therefore it's a purely imaginary number.}   
\label{fig:FullContour}
\end{figure}
The partition function $Z$ on the Keldysh contour is given by
\begin{equation}
 \frac{Z}{Z_0} \hspace{-2pt}=\hspace{-3pt} \sum \limits_{n=0}^\infty \hspace{-2pt}\frac{(-i)^n}{n!} \hspace{-3pt}\int \limits _C \hspace{-4pt}dz_1 \dotsi \hspace{-2pt}\int \limits_C \hspace{-4pt}dz_n \hspace{-1pt}\langle \mathbbm{T}_C H_U(\hspace{-0.5pt}z_1\hspace{-0.5pt}) H_U(\hspace{-0.5pt}z_2\hspace{-0.5pt}) .... H_U(\hspace{-0.5pt}z_n\hspace{-0.5pt}) \rangle_0
 \label{partitionfuncqmc}
\end{equation}
where $\mathbbm{T}_C$ orders the contour times along the Keldysh contour $C$ (depicted in \autoref{fig:FullContour}), $Z_0 = \mathrm{Tr} e^{-\beta H_0}$ and
the Hubbard interaction $H_U(z)$ in the interaction picture.
Similar as in Ref. \cite{AssaadLang2007} we introduce an additional Ising spin $s^i$ into $H_U$, hence
\begin{equation}
H_U = \frac{U}{2}\sum \limits_i \sum \limits_{s^i = \pm 1}(n_{i, \uparrow} -\frac{1}{2} - s^i\delta)(n_{i, \downarrow} -\frac{1}{2} + s^i\delta)
\label{eq:HubbardU}
\end{equation}
as well as the new parameter $\delta$.
From thermodynamic QMC it is known that $\delta$ can be used to reduce the sign-problem of the simulation and we can confirm that it can still be used to tune the sign in the fully time-dependent setting.
The usual choice for 1D Hubbard models to eliminate the sign-problem is $\delta = \frac{1}{2} + 0^+$. 
Introducing that into the general expansion for the partition function \autoref{partitionfuncqmc} gives
\begin{widetext}
\begin{equation}
 \frac{Z}{Z_0} = \sum \limits_{n=0}^\infty \frac{\left( \frac{-iU}{2} \right)^n}{n!} \int \limits _C dz_1 \sum \limits_{i_1,s^{1}} \dotsi \int \limits_C dz_n \sum \limits_{i_n,s^{n}} \prod \limits_\sigma 
                 \langle \mathbbm{T}_C (n_{i_1,\sigma}(z_1) - \alpha_{\sigma,s^1})\dots (n_{i_n,\sigma}(z_n) - \alpha_{\sigma,s^n})\rangle_0
                 \vspace{-1pt}
 \label{eq:largeZ}
\end{equation}
\end{widetext}
where we introduced
 $\alpha_{\sigma,s^i} = \frac{1}{2} + \sigma s^i \delta$
and made use of the fact that for $S^z$-conserving problems the weight splits up in an $\uparrow$-part and a $\downarrow$-part.
We can compactify \autoref{eq:largeZ} by introducing \emph{configurations}.
A configuration $C_n$ consists of Hubbard vertices $V_j = [i_j, z_j, s^j]$ with their Ising spin $s^j$, hence
\begin{equation}
 C_n = \left\{ [i_1, z_1, s^1], \dots, [i_n, z_n, s^n] \right\}.
 \label{eq:Configuration}
\end{equation}
With that concept we can introduce the sum over the configuration space
\begin{equation}
 \sum \limits_{C_n} = \sum \limits_{n=0}^{\infty} \frac{1}{n!}\int \limits_Cdz_1 \sum \limits_{i_1, s^1} \dotsi \int \limits_Cdz_n \sum \limits_{i_n, s^n}.
 \label{eq:sumConfigurations}
\end{equation}
Using this notation and applying Wick's theorem to the thermal average, \autoref{eq:largeZ} can be rewritten as a sum over determinants
\begin{equation}
 \frac{Z}{Z_0} = \sum \limits_{C_n} \left( -i\frac{U}{2} \right)^n \prod \limits_{\sigma} \det(M^\sigma(C_n))
 \label{eq:nicepartitionfunc}
\end{equation}
where 
\begin{equation}
\begin{split}
&\det(M^\sigma(C_n)) = \\
&\begin{vmatrix}
 G^0_{i_1, i_1}(z_1, z_1) - \alpha_{\sigma,s^1} &  \cdots & G^0_{i_1, i_n}(z_1, z_n) \\
 \vdots & & \vdots\\
 G^0_{i_n, i_1}(z_n, z_1) & \cdots & G^0_{i_n, i_n}(z_n, z_n) - \alpha_{\sigma,s^n} \\
\end{vmatrix}.
\end{split}
\end{equation}
The entries of $M^\sigma(C_n)$ are given by the free Green's function
\begin{equation}
\begin{split}
 M^\sigma(C_n)_{i,k} &= G^{0}_{i,k}(z_i, z_k) \\
 &= \langle \mathbbm{T}_C c^\dagger_i(z_i) c_k(z_k) \rangle_0 -\delta_{ik} \alpha_{\sigma, s^i}.
 \end{split}
\end{equation}
For the Monte Carlo evaluation of the contour integrals in \autoref{eq:sumConfigurations} we have to transform them to linear integrals. To achieve that we need to specify the parametrization of the contour. An obvious linear one is
\begin{equation}
 z(s) =
\begin{cases}
 s & s \in [0, t_{exp}]\\
2 t_{exp} - s & s \in (t_{exp}, 2 t_{exp} ]\\
-i(s - 2 t_{exp}) & s \in (2 t_{exp} , 2 t_{exp} + \beta].\\
\end{cases}
\label{parametrization}
\end{equation}
With these notations we can deduce the weight of a configuration from the partition function \autoref{eq:nicepartitionfunc}
\begin{equation}
 W(C_n) =  \left(\frac{-iU}{2} \right)^n \prod \limits_\sigma \det(M^\sigma(C_n)) F(C_n)
\end{equation}
where $F(C_n)$ collects the contribution from all phases in the configuration:
\begin{equation}
 F(C_n) = \prod \limits_{k = 0}^n \frac{dz(s)}{ds}\bigg|_{s = s_k}.
\end{equation}
To evaluate the sum over all configurations stochastically we can use a Markov process using the moves of adding and removing a vertex.
To write down their acceptance ratios  we need, additionally to the weights, the proposal probabilities of the moves.
The addition of a vertex is proposed with $T_{C_n \rightarrow C_{n+1}} = \frac{1}{2NL}$,
which corresponds to the selection of an Ising spin (from $\{\pm 1\}$), the choice of a site (from $N$ sites) and of a contour-time in the range from $[0, 2 t_{exp} + \beta]$.
The proposal probability to remove a vertex is $T_{C_{n+1} \rightarrow C_{n}} = \frac{1}{n+1}$ which corresponds to the selection of a vertex from $C_{n+1}$ which has $n~+~1$ vertices.
We still face one issue before we can write down the acceptance ratios for the Metropolis algorithm.
Since $G^0$ in the real-time setting is an arbitrarily complex value and the expressions for the weights have explicit imaginary units
a probabilistic interpretation is inhibited.
Consequently we use their absolute values $| W(C_n) |$ instead of $W(C_n)$,
but we have to compensate for this when measuring observables by keeping track of the phase of a configuration.
We write down the acceptance ratios of the moves with the imaginary units still intact, keeping in mind that while implementing them we have to use the absolute values:
\begin{equation}
\textstyle
 P_{C_n \rightarrow C_{n+1}} = \min \left( \frac{-i U N L F(C_{n+1}) \prod \limits_\sigma \det(M_\sigma(C_{n+1})) }{(n+1) F(C_n)\prod \limits_\sigma \det(M_\sigma(C_{n}))} , 1 \right)
\end{equation}
and
\begin{equation}
\textstyle
 P_{C_{n+1} \rightarrow C_{n}} = \min \left( \frac{(n+1) F(C_n)\prod \limits_\sigma \det(M_\sigma(C_{n})) }{-i U N L F(C_{n+1})\prod \limits_\sigma \det(M_\sigma(C_{n+1}))}, 1 \right).
\end{equation}
These two moves are usually sufficient for the ergodicity of the algorithm. See Ref. \cite{AssaadLang2007} for a discussion of the cases where this does not apply.
\subsection{Measurement of observables in CTQMC}
Having generated the Markov chain of configurations we can start to measure observables, e.g. the single particle Green's functions.
The Green's function can be measured by inserting the Green's function operator into the average \autoref{eq:average}, therefore it is necessary to evaluate the following expression while keeping track of the sign.
\begin{equation}
\begin{split}
 G_{ij}(s, s') &= \frac{Z_0}{Z} \sum \limits_{n=0}^\infty \frac{(-i)^n}{n!} \int \limits_C dz_1 \dotsi dz_n \times\\
 &\langle \mathbbm{T}_CH_U(z_1) \dots H_U(z_n) c^\dagger_i(z(s)) c_j(z(s')) \rangle_0 \\
                &\hspace{-22.5pt}= \frac{\sum \limits_{C_n}\left( - \frac{iU}{2} \right)^n F(C_n)\prod \limits_\sigma \det(M_\sigma(C_n)) \langle \langle G_{ij}(s, s') \rangle \rangle_{C_n}}
		{\sum \limits_{C_n} \left( - \frac{iU}{2} \right)^n F(C_n)\prod \limits_\sigma \det(M_\sigma(C_n))} \\
		&\hspace{-22.5pt}= \frac{\sum \limits_{C_n} W(C_n) \langle \langle G_{ij}(s, s') \rangle \rangle_{C_n}}{\sum \limits_{C_n} W(C_n)}
 \end{split}
 \label{eq:Greensfunctionmeasurement}
\end{equation}
where we have similarly to Ref. \cite{AssaadLang2007} introduced the contribution of one configuration to the observable
\begin{equation}
 \langle \langle G_{ij}(s, s') \rangle \rangle_{C_n} \hspace{-6pt}=\hspace{-2pt} \frac{\langle \mathbbm{T}_C H_U\hspace{-0.5pt}(z_1\hspace{-0.5pt})\hspace{-0.5pt} \dots \hspace{-0.5pt}H_U\hspace{-0.5pt}(\hspace{-0.5pt}z_n\hspace{-0.5pt}) c^\dagger_i\hspace{-1pt}(\hspace{-1pt}z(s)) c_j(\hspace{-1pt}z(s')) \rangle_0}
 {\langle \mathbbm{T}_C H_U(z_1) \dots H_u(z_n) \rangle_0}\hspace{-1pt}.
\end{equation}
Now we are at the right spot to elaborate a bit on the sign problem. As stated before we have to replace the true weight $W(C_n)$ by its absolute value $|W(C_n)|$.
We can repair this by rewriting the last line of \autoref{eq:Greensfunctionmeasurement} with $W(C_n) = |W(C_n)| \pi(C_n)$. 
We have introduced the phase-factor
$\pi(C_n) = e^{i \arg(W(C_n))} = \frac{W(C_n)}{|W(C_n)|}$. Then
\begin{equation}
\begin{split}
 G_{ij}(s, s') &= \frac{\sum \limits_{C_n} W(C_n) \langle \langle G_{ij}(s, s') \rangle \rangle_{C_n}}{\sum \limits_{C_n} W(C_n)} \\
               &= \frac{\sum \limits_{C_n} |W(C_n)| \pi(C_n) \langle \langle G_{ij}(s, s') \rangle \rangle_{C_n}}{\sum \limits_{C_n} |W(C_n)| \pi(C_n)}.
\end{split}
\end{equation}
Expanding this fraction by $\frac{1}{\sum \limits_{C_n} |W(C_n)|}$ gives:
\begin{equation}
 \begin{split}
G_{ij}(s, s') &= \frac{\frac{\sum \limits_{C_n} |W(C_n)| \pi(C_n) \langle \langle G_{ij}(s, s') \rangle \rangle_{C_n}}{\sum \limits_{C_n} |W(C_n)|}}{\frac{\sum \limits_{C_n} |W(C_n)| \pi(C_n)}{\sum \limits_{C_n} |W(C_n)|}} \\
	       &= \frac{\langle \pi G_{ij}(s, s') \rangle}{\langle \pi \rangle}.
\end{split}
\end{equation}
That way we see that measuring physical observables requires keeping track of the phase-afflicted observable and of the phase itself. The average value of the true physical observable is then determined as their ratio.
For the reduction of higher Green's functions to single particle Green's functions see the algebraic identities given in Ref. \cite{DavidsPaper2010}.
Equivalent methods have been published independently \cite{PhysRevB.81.085126, PhysRevB.81.035108}.
For an overview of the algorithms the reader is referred to the review article \cite{2010arXiv1012.4474G}.

\subsection{Auxiliary field quantum Monte Carlo approach}
\label{subsec:AUXQMC}
The time evolution with a single particle Hamiltonian can be very efficiently computed with the auxiliary field quantum Monte Carlo method provided  that the stochastic evaluation of the thermal density matrix does not suffer from the negative sign problem.  The starting point is the Trotter decomposition of the imaginary-time propagation 
\begin{equation}
   e^{-\beta \left( H_0 + H_U\right) } = \lim_{M \rightarrow \infty}
   \prod_{n=1}^{M} e^{-\Delta \tau  H_0/2} e^{-\Delta \tau  H_U }   e^{-\Delta \tau H_0/2}
\end{equation}
($\Delta \tau = \beta / M $)
followed by the  Hubbard-Stratonovitch decoupling of the Hubbard interaction,
\begin{equation}
   e^{-\Delta \tau H_U} = \frac{e^{N\Delta \tau U /4} }{2^N}\sum_{s_1 \cdots s_N = \pm 1} 
   e^{ \alpha \sum_{i} s_i \left( n_{i,\uparrow} -n_{i,\downarrow} \right)}
\end{equation}
with $\cosh(\alpha) = e^{\Delta \tau U/2}$.
This allows to  express the thermal density matrix -- or imaginary-time propagation -- in terms of a sum of non-interacting problems in an external field,
at the expense of introducing a systematic error.
In particular, for a given number of Trotter slices $M$,
\begin{widetext}
\begin{equation}
   e^{-\beta H } = \sum_{ \left\{ s_{i, n} \right\} }
      \underbrace{ 
      \frac {e^{N \beta U /4} }{2^{MN} } \prod_{k=1}^{M}
      e^{-\Delta \tau  H_0/2}  
      e^{ \alpha \sum_{i} s_{i,k} \left( n_{i,\uparrow} -n_{i,\downarrow} \right)}
      e^{-\Delta \tau  H_0/2}}_{\equiv U_{{\pmb s}}(\beta,0)} \;\;
       + {\cal O}(\Delta \tau ^2).
\end{equation}
\end{widetext}
Note that the Hubbard Stratonovitch field has acquired an extra imaginary-time index $k$.

Within this approach the real-time expectation value of an observable $O$ after a quench 
at time $t=0$ to a non-interacting Hamiltonian $H_0$ reads:
\begin{equation}
\begin{split}
   \langle O \rangle (t) &= \frac{\sum_{{\pmb s}} {\rm Tr} 
              \left[ U_{\pmb s}(\beta,0)   
             e^{i t H_0 } O  e^{-i t H_0}  \right] }
         {\sum_{{\pmb s}}  {\rm Tr} \left[U_{\pmb s}(\beta,0) \right]} \\
&\equiv 
\frac{\sum_{{\pmb s}} {\rm Tr} 
              \left[ U_{\pmb s}(\beta,0)      \right]  
               \langle \langle O \rangle \rangle_{\pmb s}(t)
                }
         {\sum_{{\pmb s}}  {\rm Tr} \left[U_{\pmb s}(\beta,0) \right]} 
\end{split}
\end{equation}
with 
\begin{equation}
               \langle \langle O \rangle \rangle_{\pmb s}(t)  = 
                \frac{ {\rm Tr}  \left[ U_{\pmb s}(\beta,0)   
                 e^{i t H_0 } O  e^{-i tH_0}   \right]  }
                     { {\rm Tr}  \left[ U_{\pmb s} (\beta,0) \right] }.
\end{equation}
For a fixed Hubbard Stratonovitch configuration, Wick's theorem applies such that
the knowledge of the single particle Green's function,
\begin{equation}
          \left[ G_{\pmb s}(t)\right]_{x,y}  =   \langle \langle c_x c^{\dagger}_y \rangle \rangle_{\pmb s}(t),
\end{equation}
suffices to compute the time evolution of any quantity.
For 
\begin{equation}
   H_0 = \sum_{x,y}c^{\dagger}_x  \left[ \mathcal{H}_0 \right]_{x,y} c_y
\end{equation}
the Green's function matrix satisfies the equation of motion: 
\begin{equation}
   \frac{d }{d t }G_{\pmb s} (t)  = -i \left [ \mathcal{H}_0 , G_{\pmb s} (t) \right],
\end{equation}
such that 
\begin{equation}
   G_{\pmb s} (t)  =   e^{-it \mathcal{H}_0 } G_{\pmb s} (t=0) e^{it \mathcal{H}_0 }
\end{equation}
with the Hamiltonian matrix $\mathcal{H}_0$.
The above equation reveals how to generalize standard finite temperature implementations of  the auxiliary field
algorithm to account for quenches to non-interacting Hamiltonians. In particular for each realization of  the 
Hubbard Stratonovitch field the equal time Green's function matrix, which correspond to the central quantity in the
 algorithm,  can be propagated according to the above equation at the expense of two  matrix multiplications. 
Using Wicks theorem, arbitrary correlation functions at a given time $t$  can be computed.
For an introduction to the usual imaginary-time formulation of the auxiliary field QMC see Ref. \cite{Assaad08_rev}.

\section{Application to quenched Hubbard models}
\label{sec:numresults}
\subsection{Thermalization towards a free model}
In \autoref{fig:docc1D} we show the time-resolved double occupancy.
Starting from its initial value in the Mott-insulating state
the double occupancy shoots up to a value larger than that of free electrons ($\langle n_{\uparrow}n_{\downarrow}\rangle = 0.25$)
and peaks at a time of $t \approx 0.61$ independently of the initially chosen $U$.
This coincides nicely with the first zero of $J_0(4 x)$.
Furthermore, the period of the following oscillations is independent on $U$.
This confirms our approximate analytic result that the frequency of the oscillations mostly depends on the band width.
In the {\it long time} limit $\langle n_{\uparrow}n_{\downarrow}\rangle$ approaches the non-interacting value.
(Note that we limited the plot to a maximal time of $t = 16$ as this is the time scale where the finite size
effects due to the boundary set in.)
\begin{figure}
\begin{center}
     The double occupancy $\langle n_{\uparrow}n_{\downarrow}\rangle$ for different values of U
\end{center}
\psfrag{title}[Bl][Bl][0.7]{$N=64, \beta=10$}
\psfrag{0.2}[Bl][Bl][0.7]{$0.2$}
\psfrag{0.1}[Bl][Bl][0.7]{$0.1$}
\psfrag{0.15}[Bl][Bl][0.7]{$0.15$}
\psfrag{0.25}[Bl][Bl][0.7]{$0.25$}
\psfrag{0.3}[Bl][Bl][0.7]{$0.3$}
\psfrag{t}[Bl][Bl][0.7]{$t$}
\psfrag{10}[Bl][Bl][0.7]{$10$}
\psfrag{0}[Bl][Bl][0.7]{$0$}
\psfrag{5}[Bl][Bl][0.7]{$5$}
\psfrag{15}[Bl][Bl][0.7]{$15$}
\psfrag{U=4}[Bl][Bl][0.8]{$U=4$}
\psfrag{U=3}[Bl][Bl][0.8]{$U=3$}
\psfrag{U=2}[Bl][Bl][0.8]{$U=2$}
\psfrag{U=1}[Bl][Bl][0.8]{$U=1$}
\psfrag{non-interacting case}[Bl][Bl][0.8]{non-interacting case}
\psfrag{ylabel}[Bl][Bl][0.8]{$\langle n_\uparrow n_\downarrow \rangle$}
\includegraphics[width=\linewidth]{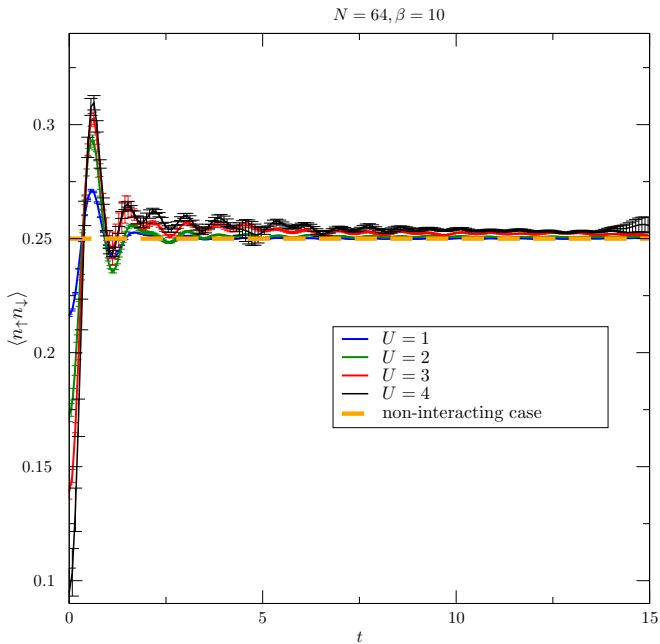}
\caption{The double occupancy of the 1D Monte Carlo simulations seems to decay to $0.25$, the value of a free model.}
\label{fig:docc1D}
\end{figure}

\begin{figure}
\caption{Spin spin correlations in 1D and 2D}
\addtocounter{figure}{-1}
\begin{tabular}{cc}
\normalsize
$S(q=\pi,t)$ &
\normalsize
 $\lvert S(q,t) - S_{\text{eff}}(q) \rvert, q = \pi$ \\
 \subfigure[\hspace{0.2cm}The spin spin correlation functions decay towards the values given by the effective model (dashed line).]
 {
 \psfrag{t}[Bl][Bl][0.7]{$t$}
\psfrag{10}[Bl][Bl][0.5]{$10$}
\psfrag{0}[Bl][Bl][0.5]{$0$}
\psfrag{5}[Bl][Bl][0.5]{$5$}
\psfrag{15}[Bl][Bl][0.5]{$15$}
\psfrag{U=4}[Bl][Bl][0.5]{$U=4$}
\psfrag{U=3}[Bl][Bl][0.5]{$U=3$}
\psfrag{U=5}[Bl][Bl][0.5]{$U=5$}
\psfrag{Sqpi}[Bl][Bl][0.5]{$S(q=\pi,t)$}
\psfrag{hypothesis}[Bl][Bl][0.5]{\hspace{-0.1cm}$S_{\text{eff}}(q=\pi)$}
\psfrag{0.2}[Bl][Bl][0.5]{$0.2$}
\psfrag{0.4}[Bl][Bl][0.5]{$0.4$}
\psfrag{0.6}[Bl][Bl][0.5]{$0.6$}
\psfrag{0.8}[Bl][Bl][0.5]{$0.8$}
 \begin{minipage}[t]{0.46\linewidth}
   \includegraphics[width=\linewidth]{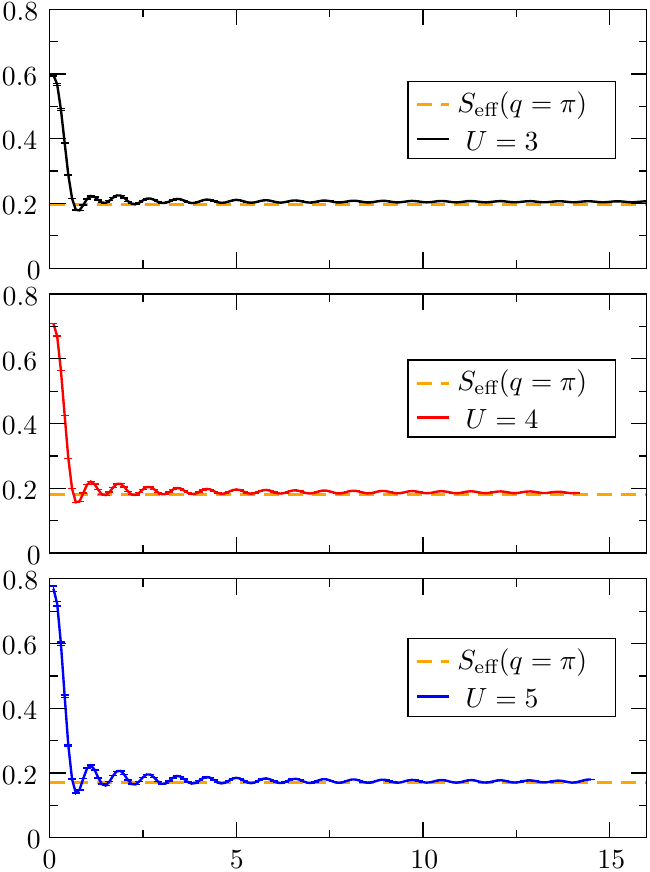}
   \vspace{-0.9cm}
   \begin{center}
    $\scriptstyle t$
   \end{center}
\end{minipage}\hfill
  \label{fig:SpinSpin1D}
 }
    &
    \subfigure[\hspace{0.2cm}In the log-log-plot we see the power-law like behaviour of $S(\pi ,t )$. The decay is roughly $t^{-1}$. Here $N=64$ and $\beta=10$.]{
\psfrag{U=4}[Bl][Bl][0.5]{$U=4$}
\psfrag{U=3}{$U=3$}
\psfrag{U=5}[Bl][Bl][0.5]{$U=5$}
\psfrag{1}[Bl][Bl][0.5]{$1$}
\psfrag{t}[Bl][Bl][0.5]{$t$}
\psfrag{10}[Bl][Bl][0.5]{$10$}
\psfrag{title}[Bl][Bl][0.5]{$N=64, \beta=10$}
\psfrag{tfitsabl}[Bl][Bl][0.5]{$\propto t^{-1}$}
\psfrag{0.1}[Bl][Bl][0.5]{$0.1$}
\psfrag{0.01}[Bl][Bl][0.5]{$0.01$}
\psfrag{0.001}[Bl][Bl][0.5]{$0.001$}
\psfrag{0.0001}[Bl][Bl][0.5]{$0.0001$}
\psfrag{1e-05}[Bl][Bl][0.5]{$10^{-5}$}
\begin{minipage}[t]{0.46\linewidth}
\includegraphics[width=\linewidth]{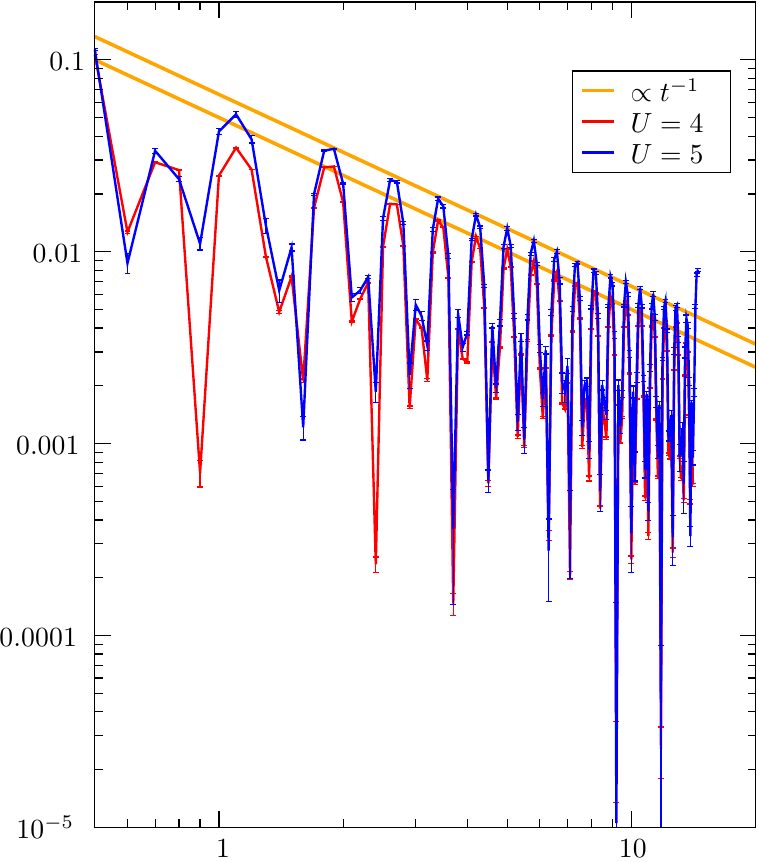}
\vspace{-0.9cm}
 \begin{center}
  $\scriptstyle t$
 \end{center}
\end{minipage}
\label{fig:SpinSpin1D_power_law}
}
\end{tabular}

\begin{tabular}{cc}
\normalsize
    $S(\vec{q}=(\pi, \pi)^T,t)$
&
\normalsize
    $\lvert S(\vec{q},t)\hspace{-1.5pt} - \hspace{-1.5pt}S_{\text{eff}}(\vec{q}) \rvert, \vec{q} =\hspace{-1pt} (\pi,\hspace{-0.8pt} \pi)^T$
    \\
    \subfigure[\hspace{0.2cm}Also in 2D we see the decay towards the effective model. Note, that in contrast to the 1D simulations the maximum time is
$t \approx 4.5$ as determined from finite-size scaling.]{
\psfrag{Ub=6}[Bl][Bl][0.5]{$U=6$}
\psfrag{Ub=4}[Bl][Bl][0.5]{$U=4$}
\psfrag{Ub=2}[Bl][Bl][0.5]{$U=2$}
\psfrag{1}[Bl][Bl][0.45]{$1$}
\psfrag{1.5}[Bl][Bl][0.45]{\hspace{-0.1cm}$1.5$}
\psfrag{2}[Bl][Bl][0.45]{$2$}
\psfrag{2.5}[Bl][Bl][0.45]{\hspace{-0.1cm}$2.5$}
\psfrag{3}[Bl][Bl][0.45]{$3$}
\psfrag{4}[Bl][Bl][0.45]{$4$}
\psfrag{t}[Bl][Bl][0.5]{$t$}
\psfrag{0}[Bl][Bl][0.45]{$0$}
\psfrag{10}[Bl][Bl][0.45]{$10$}
\psfrag{20}[Bl][Bl][0.45]{$20$}
\psfrag{30}[Bl][Bl][0.45]{$30$}
\psfrag{40}[Bl][Bl][0.45]{$40$}
\psfrag{50}[Bl][Bl][0.45]{$50$}
\psfrag{60}[Bl][Bl][0.45]{$60$}
\psfrag{title}[Bl][Bl][0.5]{$20x20$-System, $\beta=10$}
\includegraphics[width=0.46\linewidth]{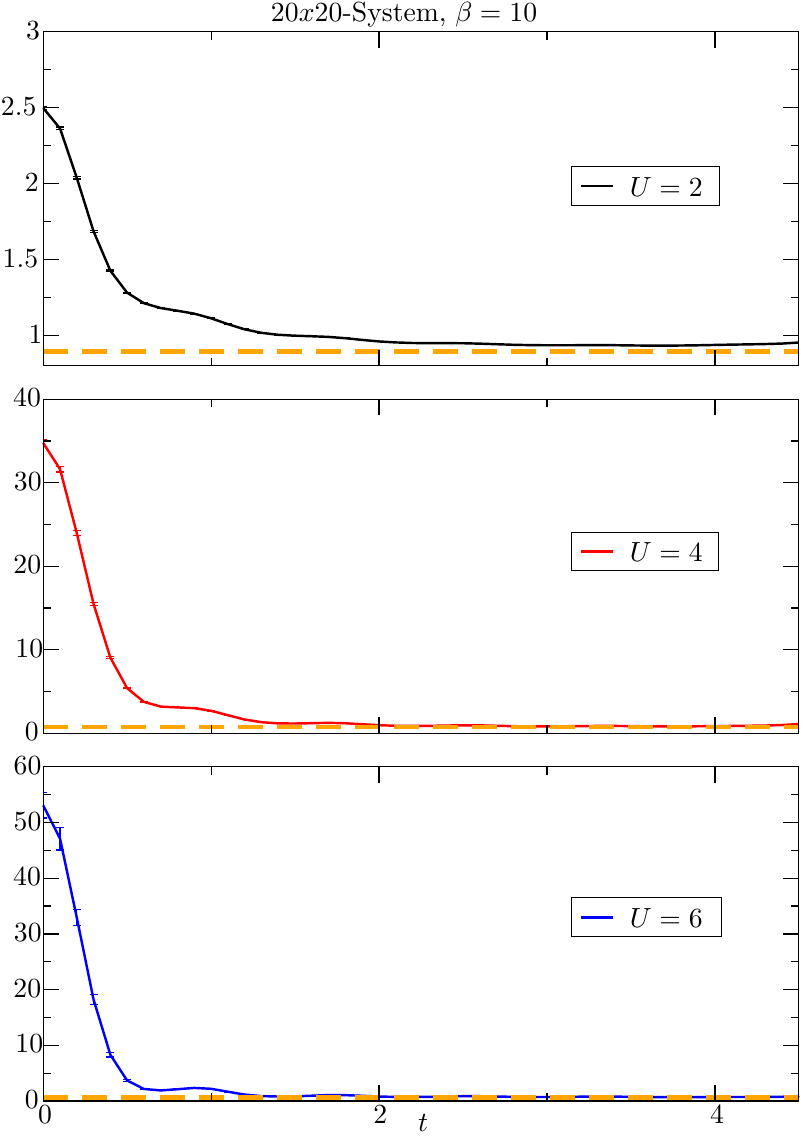}
\label{fig:SpinSpin2D}
}
&
\subfigure[\hspace{0.2cm}In 2D, the envelope of the oscillations follows a long-time decay law $\propto t^{-2}$. 
For $U=8$ and $\beta = 2.5$, this is supported by the dashed line which is $\propto J^4_0(4t)$]{
\psfrag{100}[Bl][Bl][0.5]{\hspace{-0.1cm}$100$}
\psfrag{10}[Bl][Bl][0.5]{\hspace{-0.1cm}$10$}
\psfrag{1}[Bl][Bl][0.5]{$1$}
\psfrag{2}[Bl][Bl][0.5]{$2$}
\psfrag{0.25}[Bl][Bl][0.5]{\hspace{-0.1cm}$0.25$}
\psfrag{0.5}[Bl][Bl][0.5]{\hspace{-0.1cm}$0.5$}
\psfrag{0.1}[Bl][Bl][0.5]{\hspace{-0.1cm}$0.1$}
\psfrag{0.01}[Bl][Bl][0.5]{\hspace{-0.1cm}$0.01$}
\psfrag{0.001}[Bl][Bl][0.5]{\hspace{-0.1cm}$0.001$}
\psfrag{t}[Bl][Bl][0.5]{$t$}
\psfrag{U=6}[Bl][Bl][0.5]{$U=6, \beta = 10$}
\psfrag{U=4}[Bl][Bl][0.5]{$U=4, \beta = 10$}
\psfrag{U=8}[Bl][Bl][0.5]{$U=4, \beta = 2.5$}
\psfrag{title}[Bl][Bl][0.5]{$20x20$ sites System}
\psfrag{tsquared}[Bl][Bl][0.5]{$\propto t^{-2}$}
\psfrag{Besselfunction}[Bl][Bl][0.5]{$\propto J_0^4(4t)$}
\includegraphics[width=0.46\linewidth]{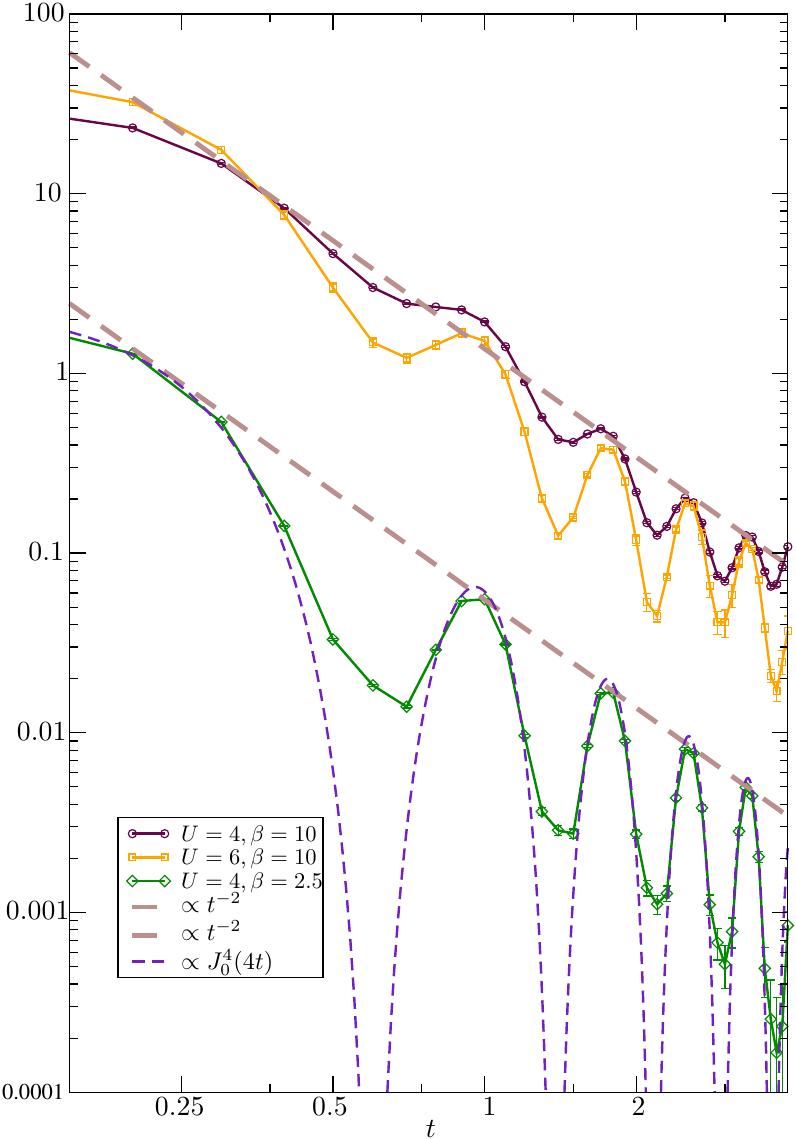}
\label{fig:SpinSpin2D_Power_Law}
}
\\
\end{tabular}
\label{fig:SpinSpinkSpace}
\addtocounter{figure}{1}
\end{figure}
Since the double occupancy equilibrates to the value of free electrons we can conjecture that the long time stationary behaviour is described
by an effective non-interacting model supplemented with Lagrange multipliers that enforce the conservation laws.
We therefore propose an effective density matrix of the form
\begin{equation}
 \rho_{\text{eff}} = \frac{e^{-\beta_{\text{eff}} H_{\text{eff}}}}{\text{Tr} e^{-\beta_{\text{eff}} H_{\text{eff}}}}
\end{equation}
where the effective Hamiltonian is of the form
\begin{equation}
 H_{\text{eff}} = \sum \limits_{ij\sigma} c^\dagger_{i\sigma} T_{ij}c_{j\sigma} + \sum \limits_{k\sigma} \lambda_{k\sigma} (n_{k\sigma} - \langle n_{k\sigma}(t=0) \rangle )
\end{equation}
with undetermined Lagrange multipliers $\lambda_{k\sigma}$ and an arbitrary single particle Hamiltonian set by $T_{ij}$.
With this ansatz one can predict uniquely the long time stationary value of any correlation function.
In particular consider the equal time spin spin correlations. Owing to Wick's theorem they are uniquely determined by the single particle occupations,
\begin{equation}
\begin{split}
 S_{\text{eff}}(q) &\propto \frac{1}{N}\sum \limits_{\substack{
                                          \sigma \sigma'\\k p
                                         }}
 \sigma \sigma' \langle c^\dagger_{k\sigma} c_{k+q, \sigma} c^\dagger_{p \sigma'}c_{p-q, \sigma'}\rangle_{\text{eff}}\\
 &\propto \frac{1}{N} \sum \limits_{\sigma k} \langle n_{k\sigma}\rangle_{\text{eff}}(1-\langle n_{k+q, \sigma}\rangle_{\text{eff}}).
 \end{split}
\end{equation}
Since the single particle occupation numbers are conserved quantities the long time behaviour of any correlation function can be uniquely determined from the knowledge of
$\langle n_{k\sigma} \rangle$ at $t=0$.
We test this prediction by computing the time-dependent spin spin correlation functions:
\begin{equation}
 S(q,t) = \frac{1}{N} \sum \limits_{\sigma\sigma'k p } \sigma \sigma' \langle c^\dagger_{k\sigma} c_{k+q, \sigma} c^\dagger_{p \sigma'}c_{p-q, \sigma'}\rangle(t).
\end{equation}
\autoref{fig:SpinSpin1D} shows the behaviour of $S(q=\pi,t)$ in 1D and \autoref{fig:SpinSpin2D} in 2D.
In both considered dimensions, the spin spin correlations decay rapidly and approach the values determined with the effective model.
Note that in 1D it takes longer for the oscillations to fade out than in 2D. This is consistent with our calculations from perturbation theory
which predict a smaller exponent in 1D than in 2D.
\subsection{Decay of correlation}
To extract the decay rate of the correlation functions we plotted in 
\autoref{fig:SpinSpin1D_power_law} and \autoref{fig:SpinSpin2D_Power_Law} the difference to the effective model on a log-log scale.
We see that the maxima of each of the oscillations can roughly be fitted by straight lines, thus the decay shows power-law behaviour.
Due to the rather low linear dimension of the 2D system the observable time evolution is restricted by finite-size effects to about $t=4.5$.
In 1D we observe a decay with a power law like $t^{-1}$ whereas in 2D as $t^{-2}$. This is consistent with
the previous analytic considerations.

To conclude we observe that the system relaxes to a state that is well described by a fermionic gaussian Hamiltonian.
At least for bosons Cramer et al \cite{CramerEisert2010}
have published proofs that the time-evolution of an arbitrary initial state
under a quadratic Bose Hamiltonian -- therefore some kind quench dynamics --
leads to local relaxation towards gaussian Hamiltonians.
Physically, the authors argue that this is due to the effect that every subsystem acts like a bath for the other, 
while their coupling is mediated by the local interactions.
\FloatBarrier
\subsection{Information propagation in correlation functions}
To study the information propagation in the system we consider two particle correlation functions.
Information propagation has already been studied for a 1D Bose-Hubbard model in Ref. \cite{LaeuchliCollath2008} and for spinless fermions in Ref. \cite{Manmana2009}.
\begin{figure}
\begin{center}
 Spatially resolved charge charge correlation functions for different times
\end{center}
\psfrag{1}[Bl][Bl][0.7]{$1$}
\psfrag{0}[Bl][Bl][0.7]{$0$}
\psfrag{0.0001}[Bl][Bl][0.7]{$0.0001$}
\psfrag{16}[Bl][Bl][0.7]{$16$}
\psfrag{32}[Bl][Bl][0.7]{$32$}
\psfrag{48}[Bl][Bl][0.7]{$48$}
\psfrag{64}[Bl][Bl][0.7]{$64$}
\psfrag{1e-08}[Bl][Bl][0.7]{$10^{-8}$}
\psfrag{1.6}[Bl][Bl][0.7]{$1.6$}
\psfrag{3.2}[Bl][Bl][0.7]{$3.2$}
\psfrag{4.8}[Bl][Bl][0.7]{$4.8$}
\psfrag{6.4}[Bl][Bl][0.7]{$6.4$}
\psfrag{8}[Bl][Bl][0.7]{$8.0$}
\psfrag{9.6}[Bl][Bl][0.7]{$9.6$}
\psfrag{11.2}[Bl][Bl][0.7]{$11.2$}
\psfrag{12.8}[Bl][Bl][0.7]{$12.8$}
\psfrag{14.4}[Bl][Bl][0.7]{$14.4$}
\includegraphics[width=\linewidth]{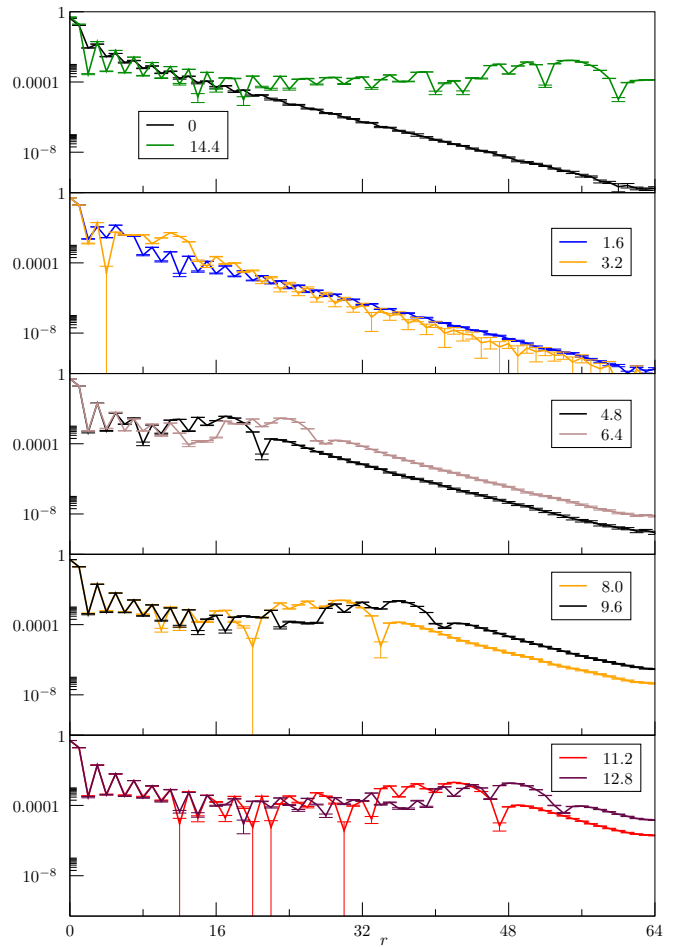}
\caption{The spatially resolved charge charge correlation: $|\langle n_0(t) n_i(t)\rangle - \langle n_0(t)\rangle \langle n_i(t)\rangle |$.
For $t=0$ we see the characteristic exponential decay of an insulator. Between $t=1.6$ and $t=3.2$ we see
that a characteristic front forms that is propagating through the lattice. The area behind this front seems to be 
metallic as evidenced by the lack of an exponential decay.
This is a lattice of 128 sites at $\beta = 10$ with an initial $U=1$.
}
\label{fig:chargechargespatial}
\end{figure}

\autoref{fig:chargechargespatial} plots the spatially resolved charge charge correlation functions as a function of time.
At $t=0$ we observe the characteristic exponential decay of this quantity as appropriate for insulating states.
As a function of time a characteristic horizon forms. Beyond this horizon the charge charge correlation functions retain their exponential decay,
whereas well within the horizon time independent correlation functions emerge.

\begin{figure}
\begin{center}
Charge charge correlation functions as a function of time for different values of $U$
\end{center}
\psfrag{U=1}[Bl][Bl][0.9]{$U=1$}
\psfrag{U=3}[Bl][Bl][0.9]{$U=3$}
\psfrag{U=6}[Bl][Bl][0.9]{$U=6$}
\psfrag{0.1}[Bl][Bl][0.9]{$0.1$}
\psfrag{0.01}[Bl][Bl][0.9]{$0.01$}
\psfrag{0.001}[Bl][Bl][0.9]{$0.001$}
\psfrag{Ccorr}[Bl][Bl][0.9]{$\lvert C_{\text{corr}}(r,t) \rvert$}
\psfrag{r=1}[Bl][Bl][0.9]{$\lvert C_{\text{corr}}(1,t) \rvert$}
\psfrag{r=3}[Bl][Bl][0.9]{$\lvert C_{\text{corr}}(3,t) \rvert$}
\psfrag{r=5}[Bl][Bl][0.9]{$\lvert C_{\text{corr}}(5,t) \rvert$}
\psfrag{Chypotetr1}[Bl][Bl][0.9]{$\lvert C_{\text{eff}}(1) \rvert$}
\psfrag{Chypotetr3}[Bl][Bl][0.9]{$\lvert C_{\text{eff}}(3) \rvert$}
\psfrag{Chypotetr5}[Bl][Bl][0.9]{$\lvert C_{\text{eff}}(5) \rvert$}
\psfrag{t}[Bl][Bl][0.9]{$t$}
\psfrag{0}[Bl][Bl][0.9]{$0$}
\psfrag{5}[Bl][Bl][0.9]{$5$}
\psfrag{10}[Bl][Bl][0.9]{$10$}
\includegraphics[width=\linewidth]{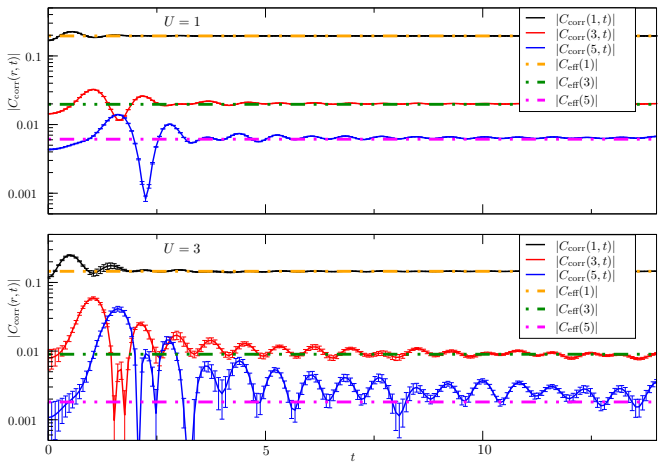}
 \caption{For a 64 site lattice we see that the time in which the correlation functions $C_\text{corr}(r, t) = \langle n_0(t) n_r(t)\rangle - \langle n_0(t)\rangle \langle n_r(t)\rangle $ equilibrate, depends
 on the initial conditions, the chosen $U$. But for $U=1$ it seems reasonable to think of the short-ranged correlation functions
 as equilibrated to the effective values $C_{\text{eff}}(r)$. Note that the $y$-axis has a logarithmic scale.
 \label{fig:equilibratedChargeCharge}
 }
 \end{figure}
To understand the nature of the decay of the charge correlations well within the horizon, we plot in \autoref{fig:equilibratedChargeCharge}
their time evolution for fixed distances $r$. As apparent, the equilibration time grows with the distance $r$ as well as with the initial value of the Hubbard interaction $U$.
However, the stationary value is consistent with our effective model such that well within the horizon, the charge charge correlation functions are given by: 
\begin{equation}
\label{eq:Charge-within-horizon}
\langle n(r) n(0) \rangle_{\text{eff}}  \propto \frac{1}{N^2} \sum \limits_{\sigma k q}  e^{i q r} \langle n_{k\sigma}\rangle(1-\langle n_{k+q, \sigma}\rangle).
\end{equation}
In the above, $\langle n_{k\sigma} \rangle $ corresponds to the single particle occupation number at time $t=0$.
Within a mean-field spin density wave approximation this quantity reads:
\begin{equation}
\label{eq:SDW_nk}
     \langle n_{k\sigma} \rangle _{SDW} = \frac{1}{2} \left( 1 - \frac{\epsilon(k)}{\sqrt{\epsilon(k)^2 + \Delta^2}}\right). 
\end{equation}
Inserting this form in Eq. \ref{eq:Charge-within-horizon} yields an exponential decay of the charge correlations.
We note that this exponential decay of the QMC data may be very well reproduced by the above equations with $\Delta \approx 0.075$.
At our largest time, $t = 14.4$ in \autoref{fig:equilibratedChargeCharge}, the  charge correlations are converged in the region $r < 16$ 
and the maximum of each oscillation is consistent with an exponential decay.
In terms of the effective model, acquaint to describing the long time stationary state, the SDW result of \autoref{eq:SDW_nk} implies that:
\begin{equation}
	\text{Tr}  \left[  \rho_{\text{eff}}  n_{k\sigma} \right] =  \langle n_{k\sigma} \rangle _{SDW}=\frac{1}{1 + e^{\beta_{\text{eff}} \epsilon_{\text{eff}}(k)}}.
\end{equation}
The  last equation defines an effective band structure as well as an effective temperature
\footnote{The effective temperature is well defined provided that the overall band width of the effective dispersion relation is fixed,
for instance, to that of the non-interacting Hamiltonian.}.
With this construction, the  state after the quench may be perceived as a metallic state at finite temperature.

Having discussed the velocity of the information propagation we get a rather clear-cut 
estimate of the time scale at which  finite-size effects set in.
In our simulations on lattices of $128$ sites the finite-size effects set in at $t \approx 16$, because the velocity of the information is $v\approx 4$ and due to the periodic boundary conditions we can effectively only use half of the lattice.
\begin{figure*}
\caption{Comparison of the charge charge and spin spin correlation functions}
\addtocounter{figure}{-1}
    \begin{tabular}{cc}
\subfigure[We see the causality cone in the charge charge correlation function. \autoref{fig:exp_suppression} corresponds to a vertical cut along the $t$-axis.]{
\includegraphics[width=0.47\linewidth]{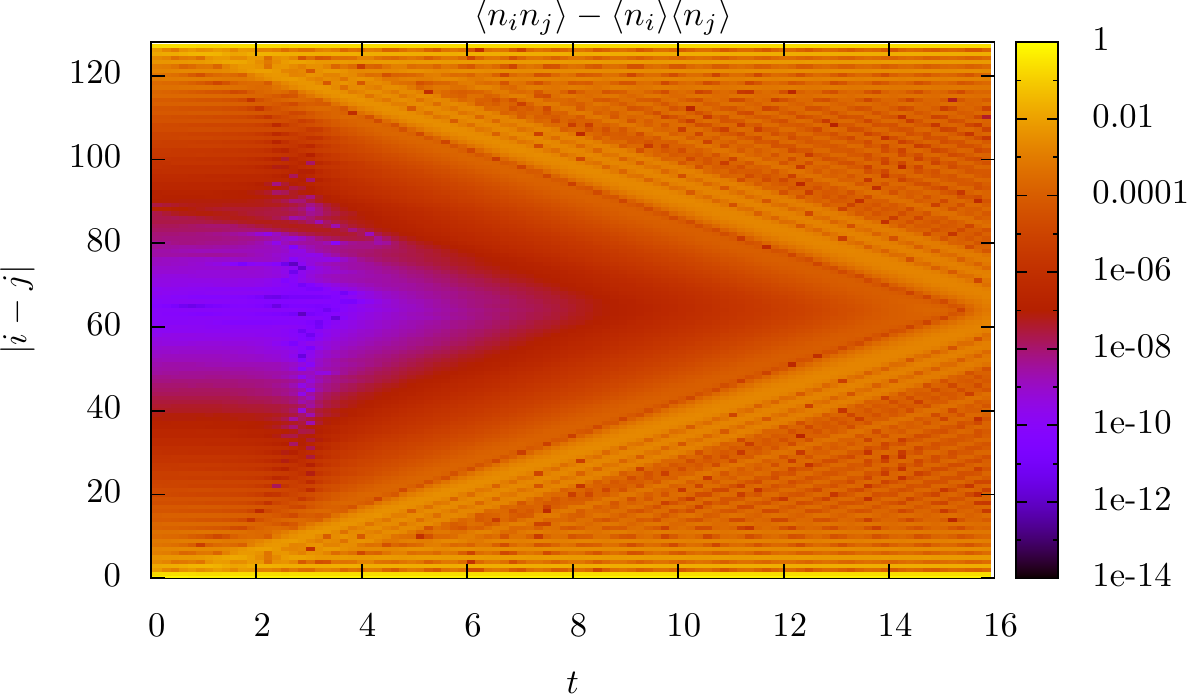}
\label{fig:ChargeChargeU1Horizon}
}
    &
    \subfigure[The spin spin correlation functions also show the horizon, but also more noise. This plot as well as \autoref{fig:ChargeChargeU1Horizon} is for a 128 site lattice at $U=1$ and $\beta = 10$.]{
\includegraphics[width=0.47\linewidth]{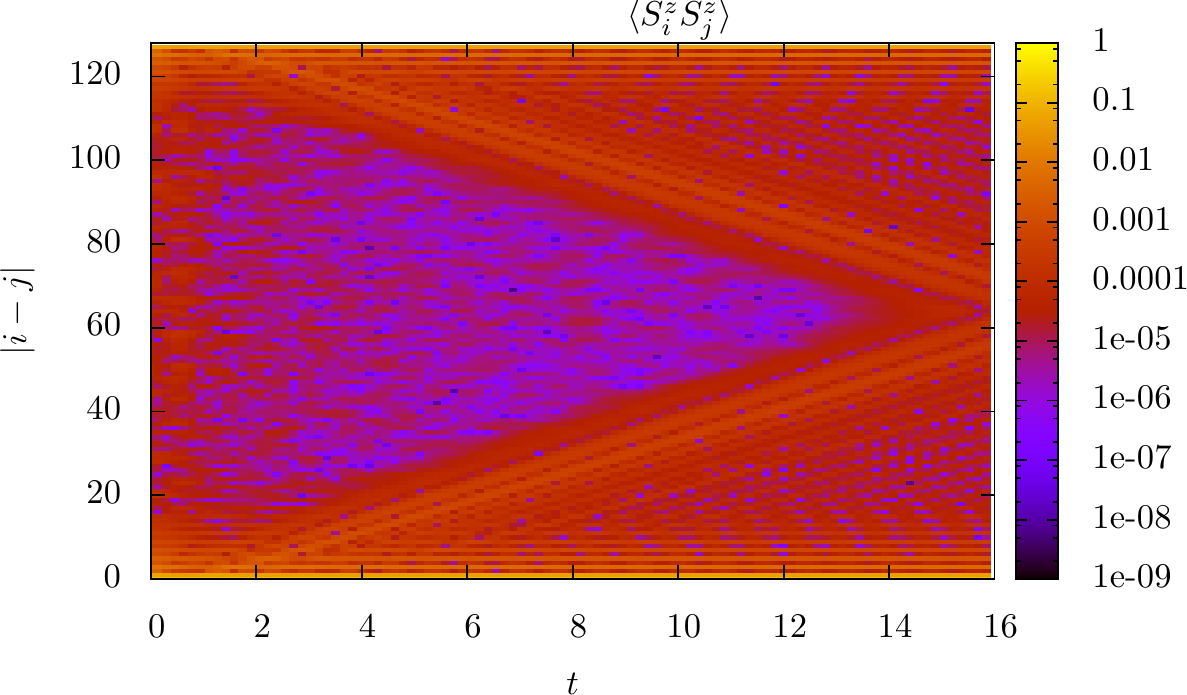}
\label{fig:SpinSpinU1horizon}
}
\end{tabular}
\end{figure*}
The torus topology of the lattice can be observed in \autoref{fig:ChargeChargeU1Horizon} and \autoref{fig:SpinSpinU1horizon} where the horizon is symmetrically expanding from the top and the bottom of the figure.
Calabrese and Cardy \cite{CalabreseCardy2007, PhysRevLett.96.136801} have put forward the picture that this information transport happens mainly by ballistic transport of 
the electrons. As we have a characteristic upper limit of the speed of the information propagation, this limit can be identified with a Lieb-Robinson bound.
Lieb-Robinson bounds are the upper limits to the group velocities of excitations traveling through the considered system. 
As already mentioned, they define a light cone like structure that gives rise to a notion of causality, since outside of the cone any influence of an excitation
is exponentially suppressed. Any non-negligible information transport is therefore limited by this speed.
To assess if we really fulfill this characteristic exponential suppression of information outside the light cone we consider some specific values of the charge charge correlation functions as a function of time.
In \autoref{fig:exp_suppression} we see that especially the longer range correlation functions show an exponential build-up of correlation outside of the causality cone.

\begin{figure}
\addtocounter{figure}{+1}
\begin{center}
 Exponential suppression outside of the causality cone of the charge charge correlation
\end{center}
\vspace{0.2cm}
\psfrag{0.1}[Bl][Bl][0.9]{$0.1$}
\psfrag{1e-12}[Bl][Bl][0.9]{$10^{-12}$}
\psfrag{nnnnn=8}[Bl][Bl][0.9]{$n=8$}
\psfrag{n=24}[Bl][Bl][0.9]{$n=24$}
\psfrag{n=32}[Bl][Bl][0.9]{$n=32$}
\psfrag{n=40}[Bl][Bl][0.9]{$n=40$}
\psfrag{n=56}[Bl][Bl][0.9]{$n=56$}
\psfrag{Chypotetr1}[Bl][Bl][0.9]{$\lvert C_{\text{eff}}(1) \rvert$}
\psfrag{Chypotetr3}[Bl][Bl][0.9]{$\lvert C_{\text{eff}}(3) \rvert$}
\psfrag{Chypotetr5}[Bl][Bl][0.9]{$\lvert C_{\text{eff}}(5) \rvert$}
\psfrag{t}[Bl][Bl][0.9]{$t$}
\psfrag{0}[Bl][Bl][0.9]{$0$}
\psfrag{5}[Bl][Bl][0.9]{$5$}
\psfrag{10}[Bl][Bl][0.9]{$10$}
\psfrag{15}[Bl][Bl][0.9]{$15$}
\psfrag{20}[Bl][Bl][0.9]{$20$}
\includegraphics[width=\linewidth]{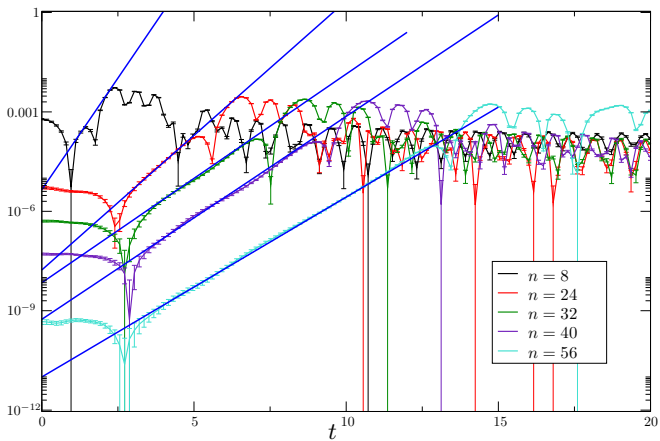}
\begin{center}
\vspace{-0.5cm}
 $t$
\end{center}

 \caption{
The exponential suppression of $|\langle n_0(t) n_i(t)\rangle - \langle n_0(t)\rangle \langle n_i(t)\rangle |$ outside of the causality cone. The blue lines are exponentials meant as a guide to the eye. The indices $n$
correspond to different distances in the measurement of the correlation function.
 }
 \label{fig:exp_suppression}
\end{figure}
Since we see a maximum velocity of information propagation as well as the exponential suppression outside of the causality cone,
we believe to have truly found the Lieb-Robinson bound in the charge sector. 
The spin-sector also shows a characteristic velocity and the exponential suppression, but our data is way more noisy for the spin spin correlations as is visible from \autoref{fig:SpinSpinU1horizon}. 
The fact that we could observe this finite propagation of information is due to our ability to do lattice simulations on very long chains (at least in 1D),
in contrast to the simulation of an effective impurity-like  model as e.g. in the DMFT approximation.
Surprisingly little work has been done to include the light cone into numerical approximation schemes, although it is a characteristic feature of lattice models in non-relativistic quantum mechanics.
One exception known to the authors is Ref. \cite{2011arXiv1104.1643E}.

\section{Summary}
We have used two QMC methods which allow to tackle the general problem of quenches from correlated thermal initial states to arbitrary  one-particle Hamiltonians. 
Provided that the initial density matrix can be generated without encountering a negative sign problem,
the real-time dynamics  does not suffer from a dynamical sign problem.
This allows to access large lattices and long propagation times.
The algorithms used are generalizations of the weak coupling continuous time and auxiliary field QMC algorithms to the Matsubara-Keldysh contour. \autoref{fig:FullContour}.
As a first step we have studied the dynamical transition from a Mott insulating state to a Fermi liquid corresponding to the quench from a finite to a vanishing Hubbard repulsion $U$, both in one and two dimensions.  
We find that spin spin correlation functions and double occupancy decay towards values that can be reproduced by an effective 
single particle Hamiltonian where the particle densities $n_{k\sigma}$ are restricted to the values of the thermal density matrix.
We observe in \autoref{fig:SpinSpinkSpace} that the decay of the magnetic order depends on the dimensionality of the system.
In 1D we observe a decay where the oscillations are enveloped by a decay like $t^{-1}$ and in 2D as $t^{-2}$.
Monte-Carlo methods have proven to be reliable tools for tackling this problem, since especially 
in 2D there is a lack of non-equilibrium approximation schemes that could provide insight into the system at hand.
One exception known to the authors is an extension of Cluster Perturbation theory in Ref. \cite{PhysRevB.83.195132}.
We have compared successfully  our results  with mean-field  and perturbative calculations in both one and two dimensions. The very good agreement points to the fact that the quench pumps enough energy in the isolated system such that the detailed correlation induced properties of the initial system do not effect in any significant way the evolution to the stationary state. This is particularly striking in the one-dimensional case, since the mean-field approximation captures by no means the physics of the initial Mott insulating state.
Due to this large amount of energy released by the quench one can argue that the  isolated system goes to a {\it high temperature} state where vertex corrections can be neglected. Hence any $n$-point correlation function can be described by a product of single particle Green's functions.
Since in $D$ dimensions the single particle Green's function of a non-interacting system exhibits a diffusive envelope, $t^{-\frac{D}{2}}$,  it follows 
that an $n$-point correlation function has a long time behavior $\propto t^{-\frac{D n}{2}}$. This is confirmed by the QMC simulations both in one and two dimensions.
In the charge charge correlation functions, \autoref{fig:ChargeChargeU1Horizon}, we observe that the information propagates with a velocity of $v \approx 4$
through the lattice. This behaviour is the same as predicted by Lieb-Robinson theorems for various systems.
Thus in the charge charge correlation function, this system preserves a sense of locality. The same applies for the spin spin correlations, \autoref{fig:SpinSpinU1horizon}.
For distances within the light cone, the charge charge correlations are consistent with a {\it  power-law} decay. Beyond this length scale they follow an exponential law characteristic of the insulating state.

\section{Acknowledgments}
We especially thank D. Luitz for helpful discussions as well as M. Bercx for proof-reading the article.
We acknowledge support from DFG Grant No.~AS120/4-3.
We thank the LRZ Munich and the J\"{u}lich Supercomputing
Centre for generous allocation of CPU time.
\bibliographystyle{utcaps}
\bibliography{bibliography}

\appendix

\section{The mean-field magnetization}
\subsection{The magnetization in 1D}
\label{subsec:mag1Dasymptotic}
\begin{figure}
\begin{center}
 Comparing the asymptotic expansion to numerical mean-field data
\end{center}
 \includegraphics[width=\linewidth]{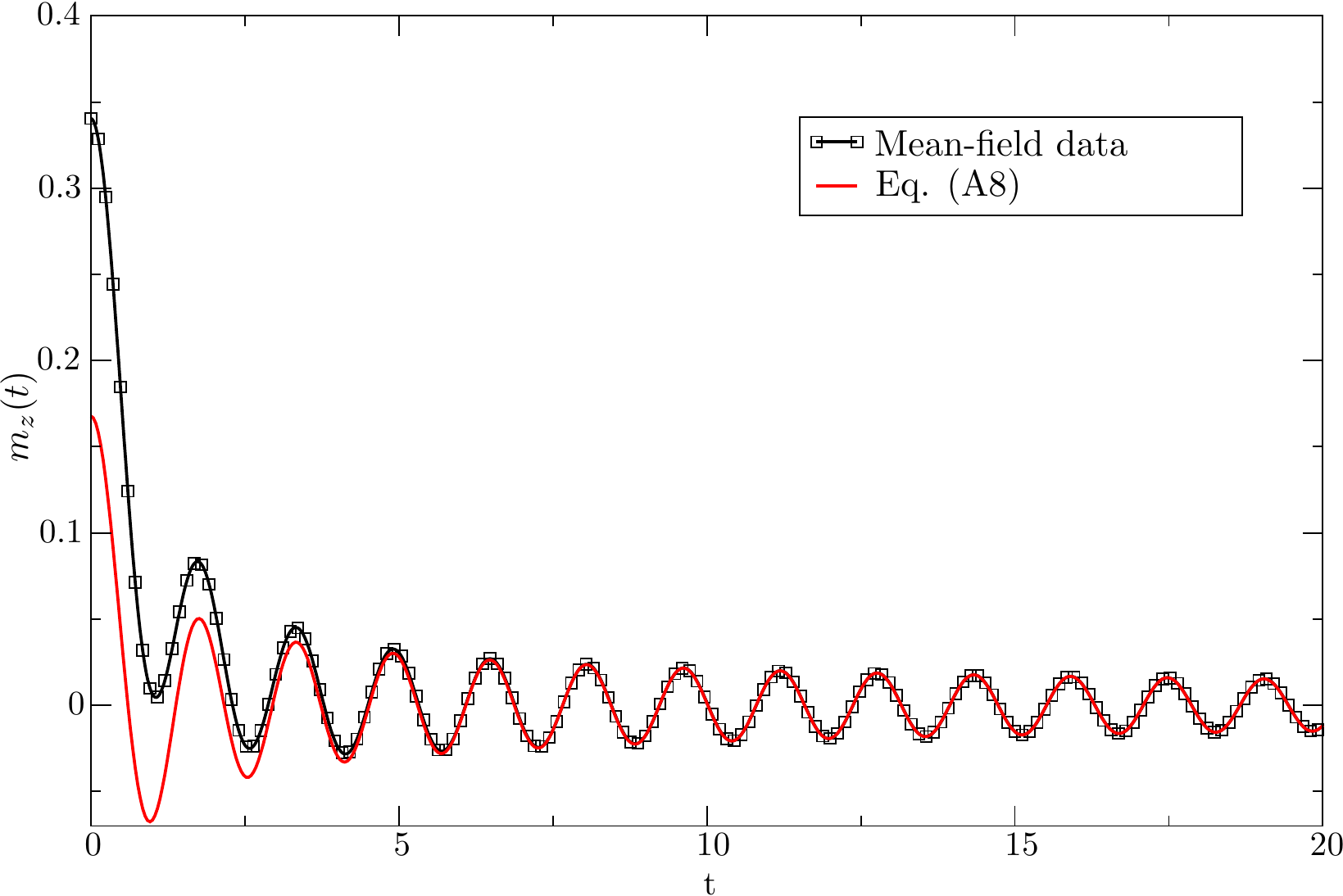}
 \caption{A comparison of numerically gained mean-field  data of a finite chain of $4096$ sites with the asymptotic expansion given in
 \autoref{eq:mz1Dasymptotic}.
 Here $U=2$ which gives a self-consistently 
 determined $\Delta \approx 0.34$.}
  \label{fig:goodasymptoticexpansion}
\end{figure}

The magnetization $m_z(\Delta, t)$ in the thermodynamic limit reads in 1D:
\begin{equation}
 m_z(\Delta, t) = \frac{2 \Delta}{\pi} \int \limits_0^2 dx \frac{1}{\sqrt{4-x^2}} \frac{1}{\sqrt{\Delta^2 + x^2}} \cos(2 t x).
 \label{eq:mz1D}
\end{equation}
Setting $t=0$ we get the initial value of the magnetization in 1D as
\begin{equation}
 m_z(\Delta, t=0) = \frac{2}{\pi}K(-\frac{4}{\Delta^2})
\end{equation}
where $K$ denotes the complete elliptic integral of the first kind.
Inserting the familiar expansion of the cosine into \autoref{eq:mz1D} and integrating term-wise we get
\begin{equation}
\begin{split}
 m_z(\Delta, t) & = \sum \limits_{k=0}^\infty (-16 t^2)^k \frac{(\frac{1}{2})_k}{(2k)! k!} \pFq{2}{1}{\frac{1}{2} + k, \frac{1}{2}}{k+1}{\frac{-4}{\Delta^2}}\\
 & = \sum \limits_{k=0}^\infty \frac{(-4 t^2)^k}{(k!)^2}  \pFq{2}{1}{\frac{1}{2} + k, \frac{1}{2}}{k+1}{\frac{-4}{\Delta^2}}
 \end{split}
\end{equation}
where $(\alpha)_k$ denotes the Pochhammer symbol and ${}_2 F_1$ is the Gauss hypergeometric function.
Using the Pfaffian transformation for the hypergeometric function we get
\begin{equation}
 m_z(\Delta,t)\hspace{-1pt}=\hspace{-1pt}\left( 1+\frac{4}{\Delta^2} \right)^{\hspace{-1pt}-\frac{1}{2}}\hspace{-1pt}
 \sum \limits_{k=0}^\infty \frac{(-4 t^2)^k}{(k!)^2} \pFq{2}{1}{\frac{1}{2}, \frac{1}{2}}{k+1}{\frac{4}{\Delta^2+4}}.
\end{equation}
Decomposing the Gauss hypergeometric function and using some properties of the Pochhammer symbol we get again a
series of hypergeometric type:
\begin{eqnarray}
 m_z(\Delta, t) &=\left( 1+\frac{4}{\Delta^2} \right)^{-\frac{1}{2}}
 \sum \limits_{k,j=0}^\infty \frac{(-4 t^2)^k}{k!} \frac{\eta^j}{j!} \frac{(\frac{1}{2})_j (\frac{1}{2})_j }{(1)_{k+j}}\\
 &= \left( 1+\frac{4}{\Delta^2} \right)^{-\frac{1}{2}} \Xi_2(\frac{1}{2}, \frac{1}{2}, 1, \eta, -4 t^2)
 \label{eq:mz1Db}
\end{eqnarray}
with the definition $\eta= 4(4+\Delta^2)^{-1}$.
$\Xi_2$ was introduced by P. Humbert to denote the twice confluent version of Appell's $F_3$ double hypergeometric function.
See Ref. \cite{Prudnikov3} chapter 7.2.4 for the definitions of hypergeometric functions in several variables.
We note the expansion of this $\Xi_2$ in terms of Bessel functions:
\begin{equation}
 \Xi_2(\frac{1}{2}, \frac{1}{2}, 1, \eta, -4 t^2)
 = 
 \sum \limits_{j=0}^\infty  \left( \frac{\eta}{2t}\right)^j \frac{(\frac{1}{2})_j (\frac{1}{2})_j}{j!} J_j(4t).
 \label{eq:mz1Da}
\end{equation}
So far, this particular $\Xi_2$ has resisted all attempts to deduce a closed form expression. 
Nevertheless it gives the asymptotic expansion with respect to $t$.
The right hand
side of \autoref{eq:mz1Da} provides a generalized asymptotic series which is asymptotic for $t\rightarrow \infty$ with respect to the asymptotic scale
$\{\phi_j \} = \{t^{-j-\frac{1}{2}}\}, j = 0,1, \dots$.
Note that for large $t$, $\Xi_2$ gets insensitive to changes in $\eta$.
This is due to the fact that the leading order behaviour of $\Xi_2$ is just $J_0(4 t)$ without any $\eta$-dependent prefactor.
To conclude, we give the leading order behaviour of $m_z(\Delta,t)$ for large $t$,
\begin{equation}
 m_z(\Delta, t) = (1+\frac{4}{\Delta^2})^{-\frac{1}{2}} J_0(4 t).
 \label{eq:mz1Dasymptotic}
\end{equation}
This function is plotted in \autoref{fig:goodasymptoticexpansion} for $U=2$ which gives $\Delta \approx 0.34$.
Obvious are the deviations for small $t$, but it is remarkable that the amplitude of the long time behaviour is very accurately described
by the inverse square root in \autoref{eq:mz1Dasymptotic}.

\subsection{The magnetization in arbitrary dimensions in the large \texorpdfstring{$\Delta$}{\textDelta} limit}
\label{subsec:magnetizationND}
We consider the mean-field magnetization on a hyper-cubic isotropic lattice of dimension D.
\begin{equation}
\begin{split}
 m_z(t,\Delta, D) & = \frac{2\Delta}{\pi}\int \limits_{-\infty}^{\infty} d\epsilon g_D(\epsilon) \frac{\cos(2 t \epsilon)}{\sqrt{\epsilon^2 + \Delta^2}}\\
                  & = \frac{2\Delta}{\pi} \operatorname{Re}{\int \limits_{-\infty}^{\infty} d\epsilon g_D(\epsilon) \frac{e^{2 t i \epsilon}}{\sqrt{\epsilon^2 + \Delta^2}}}
 \end{split}
\end{equation}
so essentially it is just the Fourier transform of some more complicated function.
But having rewritten it that way, we see that the following results also apply to the spin spin correlation
function as derived in perturbation theory \autoref{eq:Soft_small_beta}.
We start by inserting the representation of
\begin{equation}
\frac{\Delta}{\sqrt{x^2+\Delta^2}} = \InvMT{c} ds \Delta^{-s} \frac{1}{2\sqrt{\pi}} x^s \Gamma(-\frac{s}{2}) \Gamma(\frac{1+s}{2})
\label{eq:equation1}
\end{equation}
in terms of a Mellin-Barnes integral
(for properties of the Mellin transform the reader is referred to Ref. \cite{ParisAsymptotics})
where the strip of analyticity is given by $-1 < \operatorname{Re}(s) < 0$.\pagebreak[10000]
Then
\begin{equation}
\begin{split}
 &m_z(t, \Delta, D) = \frac{\Delta}{\pi}\int \limits_{-\infty}^{\infty} d\epsilon g_D(\epsilon) \frac{e^{2 t i \epsilon}}{\sqrt{\epsilon^2 + \Delta^2}}\\
 & = \frac{1}{2\pi^{\frac{3}{2}}}\InvMT{c} ds \Delta^{-s} \Gamma(-\frac{s}{2}) \Gamma(\frac{1}{2} + \frac{s}{2})\int \limits_{-\infty}^{\infty} d\epsilon g_D(\epsilon) e^{2 i t \epsilon} \epsilon^{s}.\\
 \end{split}
 \end{equation}
 Inserting the definition of the density of states this lengthens to
 \begin{widetext}
 \begin{equation}
 m_z(t, \Delta, D) = \frac{1}{2\pi^{\frac{3}{2}}}\InvMT{c} ds \Delta^{-s} \Gamma(-\frac{s}{2}) \Gamma(\frac{1}{2} + \frac{s}{2})\int \limits_{-\infty}^{\infty} d\epsilon \int dk^D \delta(\epsilon - 2\sum \limits_{i = 1}^D \cos(k_i)) e^{2 i t \epsilon} \epsilon^{s}.
 \end{equation}
Substituting $x_i=\cos(k_i)$ we get
\begin{equation}
\begin{split}
 m_z(t, \Delta, D)& = \frac{1}{2\pi^{\frac{3}{2}}}\InvMT{c} ds \Delta^{-s} \Gamma(-\frac{s}{2}) \Gamma(\frac{1}{2} + \frac{s}{2})\int \limits_{-\infty}^{\infty} d\epsilon \int_{\mathbb{R}^D} dx^D \prod_{i=1}^D \left( \frac{\Theta(1-x_i^2)}{\sqrt{1-x_i^2}} \right)
 \delta(\epsilon - 2\sum_{i=1}^D x_i) e^{2 i t \epsilon} \epsilon^{s}\\
 & = \frac{1}{2\pi^{\frac{3}{2}}}\InvMT{c} ds \Delta^{-s} \Gamma(-\frac{s}{2}) \Gamma(\frac{1}{2} + \frac{s}{2})\int_{\mathbb{R}^D} dx^D \prod_{i=1}^D \left( \frac{\Theta(1-x_i^2)}{\sqrt{1-x_i^2}} \right) e^{4 i t \sum \limits_{i=1}^D x_i} (2\sum \limits_{i=1}^D x_i)^{s}.
 \end{split}
 \end{equation}
 Displacement of the integration path to the right yields
 \begin{equation}
\begin{split}
 m_z(t, \Delta, D) & \propto \frac{1}{\pi^{\frac{3}{2}}} \sum \limits_{k=0}^M \frac{\Gamma(\frac{1}{2} + k)}{k!(-1)^k}
 \int_{\mathbb{R}^D} dx^D \prod_{i=1}^D \left( \frac{\Theta(1-x_i^2)}{\sqrt{1-x_i^2}} \right) e^{4 i t \sum \limits_{i=1}^D x_i} (2\sum \limits_{i=1}^D x_i)^{2k}\left( \frac{1}{\Delta}\right)^{2k}\\
 & = \frac{1}{\pi^{\frac{3}{2}}} \sum \limits_{k=0}^M \frac{\Gamma(\frac{1}{2} + k)}{k!4^k}
 \frac{d^{2k}}{dt^{2k}}
 \left[  \int \limits_{-\infty}^{\infty} dx \frac{\Theta(1-x_i^2)}{\sqrt{1-x_i^2}} e^{4 i t x} \right]^D
 \left( \frac{1}{\Delta}\right)^{2k} \\
 \end{split}
\end{equation}
\end{widetext}
which gives the final result
\begin{equation}
m_z(t, \Delta, D) \propto \pi^{D-1} \sum \limits_{k=0}^M \frac{(\frac{1}{2})_k}{k! (4\Delta^2)^k}
 \frac{d^{2k}}{dt^{2k}} J_0^D(4t).
 \label{eq:mzND}
\end{equation}
where we have introduced the cut-off index $M$ to terminate the asymptotic series. The $\Delta \rightarrow \infty$ limit can also be inserted 
into the high-temperature expression of the spin spin correlation function.

\subsection{The magnetization in the large \texorpdfstring{$\Delta$}{\textDelta} limit for a constant density of states}
\label{subsec:mzconst}
Assuming a constant density of states
\begin{equation}
g(\epsilon) = \frac{1}{2w}\Theta(\epsilon^2 - w^2)
\end{equation}
with bandwidth $w$ the relevant expression to analyze is
\begin{equation}
 m_z(w, \Delta, t) = \frac{\Delta}{w\pi} \int \limits_0^w dx \frac{\cos(2 t x)}{ \sqrt{\Delta^2 + x^2}}.
\end{equation}
We again perform an asymptotic expansion with respect to $\Delta$ by using \autoref{eq:equation1}. Then
\begin{equation}
 \begin{split}
 &m_z(w, \Delta, t)\hspace{-0.5pt} =\hspace{-1.5pt} \frac{1}{2w\pi^\frac{3}{2}} \hspace{-2pt}\InvMT{c}ds \Gamma(-\frac{s}{2}) \Gamma(\frac{1+s}{2}) \hspace{-1pt}\int \limits_0^w \hspace{-4pt}dx \cos(2 t x) x^s \\
 &=\hspace{-6.5pt} \InvMT{c}ds \Gamma(-\frac{s}{2}) \Gamma(\frac{1+s}{2}) \frac{w^{1+s}\Delta^{-s}}{2 w \pi^\frac{3}{2}(1+s)} \pFq{1}{2}{\frac{1}{2} + \frac{s}{2}}{\frac{1}{2}, \frac{3}{2} + \frac{s}{2}}{-t^2w^2}.
 \end{split}
\end{equation}
The poles in the positive complex half-plane are at zero and at even integers. Therefore displacement over the first $M$ poles at $s_k=2k$ yields:
\begin{equation}
 m_z(w, \Delta, t)\hspace{-2pt} \propto \hspace{-1.0pt}\frac{1}{2\pi} \sum \limits_{k=0}^{M} \hspace{-1.0pt}\frac{(-1)^k (\frac{1}{2})_k w^{2k}}{k! (1+2k)\Delta^{2k}}
 \pFq{1}{2}{\frac{1}{2} + k}{\frac{1}{2}, \frac{3}{2} + k}{-t^2w^2}\hspace{-1.5pt}.
 \label{eq:mzconst} 
\end{equation}
The first term is the contribution at infinity, therefore the $\Delta \rightarrow \infty$ limit. We get
\begin{equation}
 m_z(w, \Delta \rightarrow \infty, t) = \frac{\sin(2 w t)}{4 w t}.
\end{equation}
We proceed to evaluate the hypergeometric function by using an integral representation (\cite{Prudnikov3} Eq. 7.2.3.9).
The hypergeometric function in \autoref{eq:mzconst} is related to the spherical Bessel functions and therefore a reduction to a finite sum of elementary functions is possible. 
 Setting $x=w t$ we get
\begin{equation}
 \begin{split}
  \pFq{1}{2}{\frac{1}{2} + k}{\frac{1}{2}, \frac{3}{2} + k}{-x^2} &= \frac{\Gamma(\frac{3}{2} + k)}{\Gamma(\frac{1}{2} + k)} \int \limits_0^1 dy \, y^{k-\frac{1}{2}} \pFq{0}{1}{}{\frac{1}{2}}{-x^2 y} \\
  & = \frac{2\Gamma(\frac{3}{2} + k)}{\Gamma(\frac{1}{2} + k)} \int \limits_0^1 dz\, z^{2 k} \cos(2 x z)
 \end{split}
\end{equation}
where we have performed a variable substitution in the last line and reduced the hypergeometric function in the integrand to a cosine.
Performing integration by parts in the integral $2k$ times we get
\begin{equation}
\begin{split}
 &\int \limits_0^1 dz\, z^{2k} \cos(2 w t z) = \frac{(2 k)!}{2wt}(-(2wt)^2)^{-k} \sin(2wt) \\
 &+ \frac{1}{2wt} \sum \limits_{j=0}^{2 k - 1} (2k-j+1)_j (-2 w t )^{-j} \sin(2 w t - \frac{\pi j}{2}).
 \end{split}
\end{equation}
From this expression we immediately see that the only occurring frequency is $2 w$ and that the slowest decay is $\frac{1}{2w t }$ for all valid values
of $k$. As a by-product we have reduced the hypergeometric function in \autoref{eq:mzconst} to a finite series of elementary functions and therefore see,
that the oscillations are due to pure sines.

\end{document}